\documentclass[12pt,a4paper]{article}

\usepackage[a4paper, top = 3cm, bottom = 2.5cm, left = 2cm, right = 2cm]{geometry}
\usepackage[english]{babel}
\usepackage{color}
\usepackage[dvipsnames]{xcolor}
\usepackage{float}
\usepackage{graphicx} 
\usepackage{fancyhdr}
\usepackage{etoolbox}
\usepackage[colorlinks=true,allcolors=black]{hyperref}
\usepackage{orcidlink}

\usepackage{titlesec}
\titleformat*{\section}{\Huge\bfseries}

\usepackage{fancyhdr}
\usepackage{extramarks}

\usepackage{amsmath}
\usepackage{amssymb}
\usepackage{amsfonts}
\usepackage{amsthm}
\usepackage{cite}

\usepackage{slashed}
\usepackage{indentfirst}

\usepackage{tikz}
\usepackage{tikz-feynman}
\tikzfeynmanset{compat=1.1.0}
\usetikzlibrary{decorations.pathmorphing}  
\tikzset{
    photon/.style={
            decorate,
            decoration={snake, amplitude=1pt, segment length=5pt},
            thick
        },
    gluon/.style={
            decorate,
            decoration={coil, amplitude=1pt, segment length=4pt},
            thick
        }
}

\usepackage{acronym}
\usepackage{multicol}
\usepackage{xstring}
\usepackage{listings}

\allowdisplaybreaks[1] 

\title{Cosmological phase transitions:\\from particle physics to gravitational waves, semi-analytically}

\date{}
\begin{document}

\maketitle

\begin{center}
{
Silvia Pascoli$^{*,\dagger}$\footnote{\href{mailto:silvia.pascoli@unibo.it}{silvia.pascoli@unibo.it}}\orcidlink{0000-0002-2958-456X},
Salvador Rosauro-Alcaraz$^{\dagger,\Box}$\footnote{\href{mailto:salvador.rosauro@clermont.in2p3.fr}{salvador.rosauro@clermont.in2p3.fr}}\orcidlink{0000-0003-4805-5169}
and Matteo Zandi$^{*}$\footnote{\href{mailto:matteo.zandi2@studio.unibo.it}{matteo.zandi2@studio.unibo.it}}\orcidlink{0009-0003-6511-5183}
}

\vspace{0.7truecm}

{\footnotesize
$^*$ Dipartimento di Fisica e Astronomia, Universit\`a di Bologna,\\ via Irnerio 46, 40126 Bologna, Italy \\[.5ex]
$^\dagger$ INFN, Sezione di Bologna, viale Berti Pichat 6/2, 40127 Bologna, Italy\\[.5ex]
$^\Box$ Laboratoire de Physique de Clermont Auvergne (UMR 6533), CNRS/IN2P3,\\ Univ. Clermont Auvergne, 4 Av. Blaise Pascal, 63178 Aubi\`ere Cedex, France
}

\vspace*{2mm}

\end{center}

\renewcommand*{\thefootnote}{\arabic{footnote}}
\setcounter{footnote}{0}
\begin{abstract}
Motivated by the recent evidence of a stochastic gravitational wave background found by pulsar timing array experiments, we focus on one of the prime cosmological explanations, i.e. a supercooled first order phase transition. If confirmed, it would offer a unique opportunity to probe early Universe dynamics and the related physics beyond the Standard Model of particles and interactions. However, the prediction of the gravitational wave spectrum from a given particle physics scenario requires theoretically and computationally demanding methods. While several tools have been put forward to reduce uncertainties and automatize these computations, we study here the possibility to perform the full pipeline of computations semi-analytically in the $4D$ theory for a $U(1)^\prime$ conformal extension of the Standard Model, thus avoiding computationally intensive simulations. Our approach yields accurate results that can be used in phenomenological studies and allow for an efficient exploration of the connection between the particle physics models and their cosmological predictions.

\end{abstract}
\newpage
\section{Introduction}

With the advent of gravitational wave (GW) detection, a novel observational window into the Universe has been opened.
The recent evidence for a stochastic gravitational-wave background (SGWB) in the $\mathcal O$(nHz) frequency, reported by several pulsar timing array (PTA) experiments~\cite{NANOGrav:2023gor,NANOGrav:2023hde,Xu:2023wog,EPTA:2023fyk,Reardon:2023gzh,Miles:2024seg}, is particularly intriguing. Although supermassive black hole binaries could explain the signal~\cite{Rajagopal:1994zj,Wyithe_2003,Jaffe_2003}, this possibility requires strong deviations from standard astrophysical expectations~\cite{NANOGrav:2023pdq,NANOGrav:2023hfp} and remains much debated~\cite{Goncharov:2024htb}. The PTA signal could also have a cosmological origin, such as first-order phase transitions (FOPTs), domain walls, or scalar-induced GWs, among others~\cite{NANOGrav:2023hvm,Ellis:2023oxs,Li:2023bxy,Figueroa:2023zhu,Madge:2023dxc,Wu:2023hsa,Bringmann:2023opz,Gouttenoire:2023bqy,Chen:2023bms,Wang:2023bbc,Ghosh:2023aum,Addazi:2023jvg,Croon:2024mde,Winkler:2024olr,Conaci:2024tlc,Costa:2025csj,Costa:2025pew,Balan:2025uke,Goncalves:2025uwh}, with recent analysis finding a mild preference for a FOPT~\cite{Wang:2025wht}.

Given that there are no FOPTs in the Standard Model (SM) \cite{Kajantie:1996mn,Aoki:2006we}, the observation of a SGWB from such a process in the early Universe would represent a smoking gun signal for physics beyond the Standard Model (BSM).\footnote{This conclusion requires that astrophysical uncertainties in the interpretation of the signal to be sufficiently under control.} The FOPT is driven by a scalar field that induces the spontaneous symmetry breaking (SSB) of a symmetry of the theory through thermal corrections to its effective potential in the early Universe. The transition proceeds via tunneling from a metastable vacuum to a stable one,
leading to the nucleation of true-vacuum bubbles in the primordial plasma. These bubbles expand while interacting with the surrounding plasma, and eventually collide, potentially generating GWs~\cite{PhysRevD.30.272,10.1063/1.1338506,Kamionkowski:1993fg,Caprini:2024hue}. An observable SGWB compatible with the PTA signal has been generally related to classically scale-invariant models~\cite{Bringmann:2023opz,Goncalves:2025uwh,Balan:2025uke,Costa:2025csj}, where strong enough supercooling allows for the release of a large amount of latent heat, thus enhancing the resulting SGWB spectrum. Such a signal is typically explained by New Physics (NP) at the $\mathcal{O}(\mathrm{GeV})$ scale, pointing towards the existence of a light dark sector which interacts weakly with the SM~\cite{Agrawal:2021dbo,Antel:2023hkf}.

Gaining insight into the NP from the SGWB observation requires predicting the GW spectrum starting from an underlying particle physics model, in particular from the effective potential of the new scalar field. This effective potential contains non-local contributions that depend on the field-dependent masses of the particles running in the loops, as well as on the renormalization scale~\cite{Coleman:1973jx}. As a result, the tunneling rate must be obtained by numerically solving the bounce equations at each point in model parameter space. Since the allowed parameter space can be very large, a comprehensive numerical scan is often computationally prohibitive. Furthermore, state-of-the-art determinations of the temperature at which GWs are produced require additional numerically intensive, multi-dimensional integrations.

Steps towards a (semi-) analytic description of the whole chain of necessary computations have been taken in Refs.~\cite{Levi:2022bzt,Salvio:2023qgb,Salvio:2023ynn,Salvio:2023blb}. The common idea is to describe the effective potential as a polynomial up to quartic order, which allows to perform a one-parameter fit for the bounce action that can be exploited universally, instead of relying on numerical methods that need to be used explicitly in every point of the parameter space. While the extensively used high-temperature (HT) expansion indeed allows to write the effective potential as a polynomial, it is not accurate enough, as it is well-known that the convergence of the perturbative expansion is compromised at high temperatures~\cite{Dolan:1973qd,Weinberg:1974hy,Linde:1978px,Linde:1980ts}. Indeed, in the HT limit, the resummation of Daisy diagrams results unavoidable~\cite{Parwani:1991gq,Arnold:1992rz}. 

In this work, we go beyond previous analysis by studying not only the well-known HT expansion, but also the inclusion of Daisy contributions projected onto a basis of polynomials, allowing for a fully consistent description of the effective potential in the HT limit relevant for the tunneling process. 
Moreover, working in the $\overline{\mathrm{MS}}$ renormalization scheme, we consistently take into account the running of the couplings and of the effective potential given that the renormalization scale changes. In fact, the latter is related to the temperature of the plasma for small field values. While state-of-the-art computations rely on the dimensionally reduced theory at next-to-leading order (NLO)~\cite{Farakos:1994kx,Braaten:1995cm,Kajantie:1995dw,Gould:2021oba,Kierkla:2023von}, we build upon recent results from Ref.~\cite{Christiansen:2025xhv} to choose the renormalization scale for which the best agreement between the $4D$ description we adopt and the $3D$ effective theory is found. 

In the analysis of FOPTs there are several key temperatures that need to be determined. In addition to the extensively used critical and nucleation temperatures, the percolation one, which corresponds to the moment when 29\% of the Universe is already in the true vacuum phase, actually represents the most relevant one for GW production~\cite{Athron:2023xlk}. Given that its determination is more involved, the nucleation temperature is often used as a proxy for the moment of GW production, even though the two can differ considerably in supercooled FOPTs, and there are even examples in which one finds percolation but no nucleation temperature~\cite{Athron:2023mer}. Importantly, we propose a new method to compute the percolation temperature. By deriving an analytical expression for the fraction of the Universe in the false vacuum, the determination of the percolation temperature is reduced to a root-finding problem, analogous to the procedure used for the nucleation temperature. This method is applicable to a broad class of scenarios, not only the ones considered here. Taken together, the proposed approximations avoid computationally intensive simulations while yielding results precise enough for phenomenological analyses. 

This paper is organized as follows. In Section~\ref{sec:Veff_poly} we introduce the full pipeline of approximations that allow us to write the effective potential, including Daisy contributions, in terms of a quartic polynomial. Additionally, we discuss the renormalization scale dependence and our choice for its value as a function of the temperature. We devote Section~\ref{sec:tunneling} to summarizing the basics of tunneling and previous semi-analytic results for polynomial potentials. In Section~\ref{sec:temperatures} we introduce the different relevant temperatures in the FOPT evolution and our approximate analytical solution for the fraction of the Universe in the false vacuum, allowing us to find the percolation temperature. We apply these results to the study of the PTA signal in Section~\ref{sec:results}, for two different scenarios, and conclude in Section~\ref{sec:conclusions}.

\section{Effective potential
}\label{sec:Veff_poly}
Consider the most general Lagrangian, with gauge symmetry group $\mathcal{G}$ under which a complex scalar $\varphi$ is charged as well as some fermion fields $\psi_i$. The scalar potential is such that $\varphi$ develops a vacuum expectation value (vev), generating masses for gauge fields and other related particles. The Lagrangian for such a scenario can be generally written as
\begin{equation}
    \begin{split}
        \mathcal{L}=&-\frac{1}{4}F_{\mu\nu}^{\prime A}F^{\prime A \mu\nu}+\left(D_{\mu}\varphi\right)^{\dagger}\left(D^{\mu}\varphi\right)+i\bar{\psi}_i\slashed{D}\psi_i-\left(\bar{\psi}_iY_{ij}\psi_j\varphi+\mathrm{h.c.}\right)-V_{\mathrm{tree}}(\varphi)\,,
    \end{split}
    \label{eq:general_lagrangian}
\end{equation}
where $D_{\mu}$ is the covariant derivative, $F^{\prime A}_{\mu\nu}$ the gauge field strength tensor, and $Y$ is a Yukawa matrix. 

In order to study the dynamics of spontaneous symmetry breaking in the early Universe, we study the effective potential at one loop and at finite temperature~\cite{Quiros:1999jp}, which is given by
\begin{equation}
    \begin{split}
        V_{\mathrm{eff}}(\varphi,T)=V_{\mathrm{tree}}(\varphi)+V_{\mathrm{loop}}(\varphi)+V_{\mathrm{T}}(\varphi,T)+V_{\mathrm{Daisy}}(\varphi,T)\,,
    \end{split}
    \label{eq:general_effective_pot}
\end{equation}
in terms of the background field, which we denote for simplicity as $\varphi$ from here onward. Although uncertainties related to renormalization scale or gauge dependence~\cite{Croon:2020cgk,Athron:2022jyi,Gould:2023ovu,Lofgren:2023sep,Athron:2023xlk,Ekstedt:2024etx} impact the predictions for the GW spectrum in this approach, their effect on parameter reconstruction from a given GW signal has been shown to be mild~\cite{Lewicki:2024xan}. Our framework allows these dependencies to be explicitly incorporated by adjusting the polynomial coefficients we introduce in the following. We stress that these uncertainties can be mitigated by going to NLO in the dimensionally reduced theory~\cite{Ekstedt:2022bff,Schicho:2022wty,schicho:2021gca,Kierkla:2025vwp,Kierkla:2025qyz}, which is used in state-of-the-art computations of GWs.

The first term in Eq.~(\ref{eq:general_effective_pot}) corresponds to the tree-level potential, which we assume to be local and renormalizable, such that it can be written in terms of polynomial operators up to dimension four. The second one is the one-loop contribution to the effective potential at $T=0$~\cite{PhysRevD.7.1888}, while $V_{\mathrm{T}}$ corresponds to the finite temperature one~\cite{PhysRevD.9.3320}. Finally, $V_{\mathrm{Daisy}}$ includes the contribution from Daisy resummation~\cite{PhysRevD.45.2933}. This last term needs to be included in order to improve the infrared (IR) behavior of the potential at high temperatures. Working in the $\overline{\mathrm{MS}}$ renormalization scheme and the Landau gauge, we can respectively write $V_{\mathrm{loop}}$, $V_{\mathrm{T}}$, and $V_{\mathrm{Daisy}}$ as
\begin{equation}
    \begin{split}
        V_{\mathrm{loop}}(\varphi)=&\frac{1}{64\pi^2}\Bigg[\sum_b n_b m_b^4(\varphi)\left(\log{\frac{m_b^2(\varphi)}{\mu^2}}-c_b\right)
        -\sum_f n_f m_f^4(\varphi)\left(\log{\frac{m_f^2(\varphi)}{\mu^2}}-c_f\right)\Bigg]\,,\\
        V_{\mathrm{T}}(\varphi,T)=&\frac{T^4}{2\pi^2}\Bigg[\sum_b n_b J_b\left(\frac{m_b^2(\varphi)}{T^2}\right)
        - \sum_f n_f J_f\left(\frac{m_f^2(\varphi)}{T^2}\right)\Bigg]\,,\\
        V_{\mathrm{Daisy}}(\varphi,T)=&-\frac{T}{12\pi}\sum_b \bar{n}_b\left[\left(m_b^2(\varphi)+\Pi_b(T)\right)^{3/2}-m_b^3(\varphi)\right]\,,
    \end{split}
    \label{eq:contributions_Veff}
\end{equation}
where $m_{b\,(f)}(\varphi)$ denotes the field dependent mass for bosons (fermions) and $n_{b\,(f)}$ is its corresponding number of degrees of freedom. The constant $c_{b\,(f)}$ has values $c_b=5/6$ for vector bosons, and $c_b=3/2=c_f$ for scalars and fermions, while $\mu$ is the renormalization scale. Couplings entering in Eq.~(\ref{eq:contributions_Veff}) are implicitly evaluated at the scale $\mu$ through their RG evolution, which we discuss in detail in Subsection~\ref{sec:RGE}. Regarding the temperature dependent contributions to the effective potential in Eq.~(\ref{eq:contributions_Veff}), the thermal functions $J_{b\,(f)}(x)$ are defined as~\cite{Quiros:1999jp}
\begin{equation}
    \begin{split}
        J_{b\,(f)}(x)\equiv \int_0^{\infty}dy~ y^2\log{\left[1\mp\exp{\left(-\sqrt{y^2+x}\right)}\right]}\,.
    \end{split}
\end{equation}
At high temperatures or small field values, for which one expects $x = m^2_{b\,(f)}(\varphi)/T^2 \ll 1$, we can expand these thermal functions to find
\begin{equation}
    \begin{split}
        J_b(x)\simeq &-\frac{\pi^4}{45}+\frac{\pi^2}{12}x-\frac{\pi}{6}x^{3/2}-\frac{1}{32}x^2\log{\frac{x}{a_b}}\,,\\
        J_f(x)\simeq &\frac{7\pi^4}{360}-\frac{\pi^2}{24}x-\frac{1}{32}x^2\log{\frac{x}{a_f}}\,,
    \end{split}
    \label{eq:HT_thermal_func}
\end{equation}
where $a_b\equiv16\pi^2\exp{(3/2-2\gamma_E)}$, $a_f\equiv\pi^2\exp{(3/2-2\gamma_E)}$, and $\gamma_E$ is the Euler-Mascheroni constant. We will use this approximation extensively in the following. Lastly, we adopt the Arnold-Espinosa approach~\cite{Arnold:1992rz} to include the Daisy contribution by resumming the zero-modes for bosons, $V_{\mathrm{Daisy}}$, for which one needs to compute the thermal mass for bosons in the HT limit $\Pi_b(T)$, and $\bar{n}_b$ in Eq.~(\ref{eq:contributions_Veff}) corresponds to the number of degrees of freedom which acquire a thermal mass. We choose this approach given that it matches the prediction obtained at leading order from the dimensionally reduced theory~\cite{Lofgren:2023sep,Lewicki:2024xan}. One could resort to other methods, such as the Parwani approach in which all the modes are resummed~\cite{Parwani:1991gq}.

\subsection{High-temperature approximation}
Taking the HT expansion of the thermal functions from Eq.~(\ref{eq:HT_thermal_func}), we find the following general expression for the effective potential:
\begin{equation}
    \begin{split}
        V_{\mathrm{eff}}(\varphi,T)\simeq &V_{\mathrm{tree}}(\varphi)-\frac{\pi^2T^4}{90}\left(\sum_b n_b+\frac{7}{8}\sum_f n_f\right)+\frac{T^2}{24}\left(\sum_b n_b m_b^2(\varphi)+\frac{1}{2}\sum_f n_fm_f^2(\varphi)\right) \\
        &-\frac{T}{12\pi}\sum_bn_bm_b^{3}(\varphi)+\frac{1}{64\pi^2}\Bigg[\sum_b n_b m_b^4(\varphi) \left(\log{\frac{a_b T^2}{\mu^2}}-c_b\right) \\
        &- \sum_f n_f m_f^4(\varphi)\left(\log{\frac{a_f T^2}{\mu^2}}-c_f\right)\Bigg]+V_{\mathrm{Daisy}}(\varphi,T)\,.
    \end{split}
    \label{eq:explicit_HT_expansion_pot}
\end{equation}
Note that the logarithmic terms in $V_{\mathrm{loop}}$ and those from $V_{\mathrm{T}}$ in Eq.~(\ref{eq:general_effective_pot}) combine in such a way that the field dependent mass always cancels between both contributions~\cite{Kierkla:2023von}. This will motivate us to choose $\mu^2\propto T^2$ in the following in order to cancel potentially large logarithmic terms in the effective potential.

Leaving aside the Daisy contribution for the moment, it is clear from Eq.~(\ref{eq:explicit_HT_expansion_pot}) that, in the regime $m_{b\,(f)}(\varphi)/T\ll 1$ and assuming $m_{b(f)}\sim g_{b(f)}\varphi$ (this is always the case for classically scale invariant scenarios), one can always write the three first contributions to the effective potential from Eq.~(\ref{eq:general_effective_pot}) as a polynomial
\begin{equation}
    \begin{split}
	   V_{\mathrm{eff}}(\varphi,T)&\simeq V_0 + \frac{1}{2}m^2(T) \varphi^2 - \frac{1}{3}K(T)\varphi^3-\frac{\lambda(T)}{4}\varphi^4+V_{\mathrm{Daisy}}(\varphi,T)\\
       &\equiv V_{\mathrm{HT}}(\varphi,T)+V_{\mathrm{Daisy}}(\varphi,T)\,.
    \end{split}
	\label{eq:polynomial_shape_pot}
\end{equation}
In the second line of Eq.~(\ref{eq:polynomial_shape_pot}) we have labeled the part of the effective potential that is expanded in the HT limit as $V_{\mathrm{HT}}$, which will ease our subsequent discussions. The fact that the tunneling process falls naturally in the HT limit~\cite{Kierkla:2023von} i.e. the relevant field values for bubble nucleation satisfy $\varphi\ll T$ allowing one to exploit a universal expression to compute $S_3$~\cite{Levi:2022bzt}, provides the motivation to describe $V_{\mathrm{eff}}$ as a polynomial on $\varphi$ and determine semi-analytically the thermal parameters controlling the production of GWs. 

We note that, while the HT approximation just discussed can always be applied for $\varphi\ll T$ and for any scalar potential, the case in which SSB already happens at tree-level, as in the SM, is slightly more convoluted. First, the tunneling process is no longer dominated necessarily by the region near small field values $\varphi \ll T$, justifying the HT expansion. Additionally, scalar masses prevent a direct description of the effective potential in terms of polynomials even in the HT limit. We discuss some possible further approximations for this case in Appendix~\ref{app:tree-level-SSB}. 

In order to find explicit expressions for the coefficients $m^2(T)$, $K(T)$, and $\lambda(T)$, it is necessary to specify a tree-level potential and the particle content. For simplicity, we choose one realization of SSB giving rise to a GW signal compatible with the recent NANOGrav observation~\cite{Balan:2025uke,Goncalves:2025uwh,Costa:2025csj}: a classically scale-invariant model with a $U(1)^{\prime}$ gauge symmetry and no fermions. This will allow us to write down explicitly $V_{\mathrm{tree}}(\varphi)$ and show the main results of our analysis. The more general case with fermions can explicitly be found in Appendix~\ref{app:running_non-minimal}. However, our results can also be adopted for other choices of the gauge group $\mathcal{G}$. 

\subsection{$U(1)^{\prime}$ classically scale-invariant model}
The classically scale-invariant tree-level potential for the background field $\varphi$ is given by
\begin{equation}
    V_{\mathrm{tree}}(\varphi)=\frac{1}{4}\lambda \varphi^4\,,
    \label{eq:tree_level_pot}
\end{equation}
with covariant derivative $D_\mu\equiv \partial_\mu-i g' Z'_{\mu}$, where $Z'_{\mu}$ is the new $U(1)^{\prime}$ gauge boson. While this scenario was studied in Refs.~\cite{Salvio:2023blb,Salvio:2023qgb,Salvio:2023ynn} using an approach similar to the one we propose here, those works relied on truncating the HT expansion of the thermal functions in Eq.~(\ref{eq:HT_thermal_func}) before the logarithmic terms. As a consequence, additional simplifying assumptions were introduced in order to handle the resulting field-dependent logarithms, leading to a not sufficiently accurate approximation of the effective potential. Moreover, neither Daisy contributions nor the RG evolution were included in the analysis of the tunneling process.

Considering the one-loop corrections from the $Z'$ gauge boson as in Eq.~(\ref{eq:contributions_Veff}) at the renormalization scale $\mu=\mu_0$,\footnote{Given that couplings depend on the renormalization scale (see Subsection~\ref{sec:RGE}), we specify the couplings and the vev evaluated at $\mu=\mu_0$ with a subscript ``0''.} the scalar develops a non-zero vev $\varphi_0$, while the $Z'$ obtains a mass given by $m_{Z',0}^2\equiv g'^2_0 \varphi_0^2$. We neglect the contributions of the scalar field and the Goldstone boson, since in classically scale-invariant scenarios they are expected to be small~\cite{Kierkla:2023von}, although it is straightforward to include them~\cite{Christiansen:2025xhv}. It is customary to exchange the renormalization scale explicitly appearing in the one-loop contribution to the potential with the scalar vev. Doing so, one arrives at the following shape for the effective potential at $T=0$:
\begin{equation}
    V_{\mathrm{loop}}(\varphi)=\frac{3g'^4_0}{32\pi^2}\varphi^4\left(\log{\frac{\varphi}{\varphi_0}}-\frac{1}{4}\right)\,,
\end{equation}
where, at the scale $\mu^2=\mu_0^2\equiv m_{Z',0}^2$, the equality $\lambda_0=g'^4_0 / (16\pi^2)$ needs to be satisfied such that $\varphi_0$ is the absolute minimum. 

Nevertheless, motivated by the results in the HT limit from Eq.~(\ref{eq:polynomial_shape_pot}), we are interested in keeping the explicit dependence of the potential on $\mu^2$, so that in the following we consider the effective potential written as 
\begin{equation}
    \begin{split}
        V_{\mathrm{eff}}(\varphi,T)=&\frac{\lambda}{4}\varphi^4+\frac{3}{64\pi^2} m_{Z'}^4(\varphi)\left(\log{\frac{m_{Z'}^2(\varphi)}{\mu^2}}-\frac{5}{6}\right)+\frac{3T^4}{2\pi^2}J_b\left(\frac{m_{Z'}^2(\varphi)}{T^2}\right) \\
        &-\frac{T}{12\pi}\left[\left(m^2_{Z'}(\varphi)+\Pi_{Z'}(T)\right)^{3/2}-m^3_{Z'}(\varphi)\right]\,,
    \end{split}
    \label{eq:full_potential}
\end{equation}
where $m_{Z'}^2(\varphi)=g'^2\varphi^2$, and we find $\Pi_{Z'}(T)=g'^2T^2/3$. For the specific example of the $U(1)^{\prime}$ scale-invariant scenario at hand, we then obtain the following polynomial coefficients that enter in $V_{\mathrm{HT}}$ from Eq.~(\ref{eq:polynomial_shape_pot}):
\begin{equation}
	\begin{split}
		V_0 \equiv &-\frac{\pi^2 T^4}{30}\,,\quad
		m^2(T) \equiv  \frac{T^2}{4}g'^2\,,\\
		K(T) \equiv &\frac{3 T}{4\pi}g'^3\,,\quad
		\lambda(T) \equiv -\lambda -\frac{3}{16\pi^2}g'^4\left(\log{\frac{a_b T^2}{\mu^2}}-\frac{5}{6}\right)\,.
	\end{split}
	\label{eq:polynomial_params_NoDaisy}
\end{equation}

\begin{figure}
    \centering
    \includegraphics[width=0.49\linewidth]{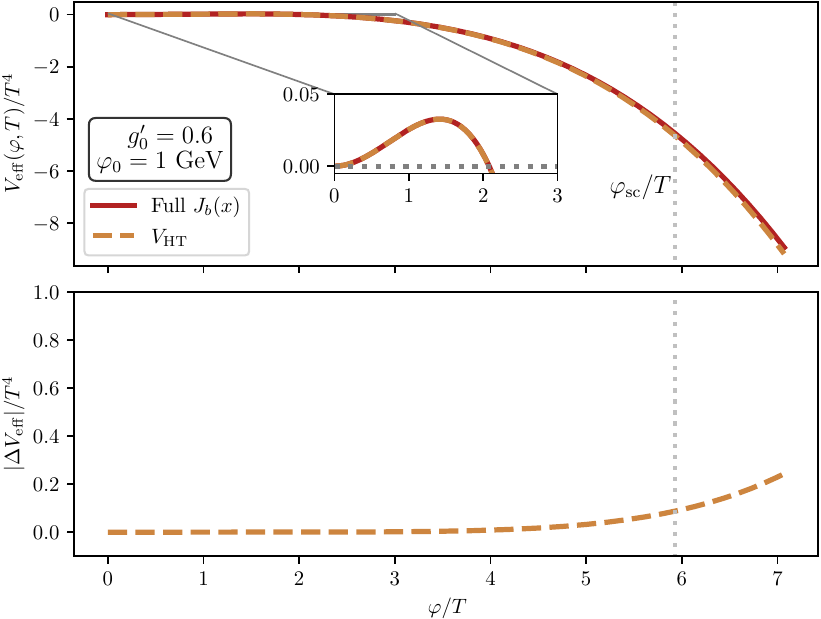}
    \includegraphics[width=0.49\linewidth]{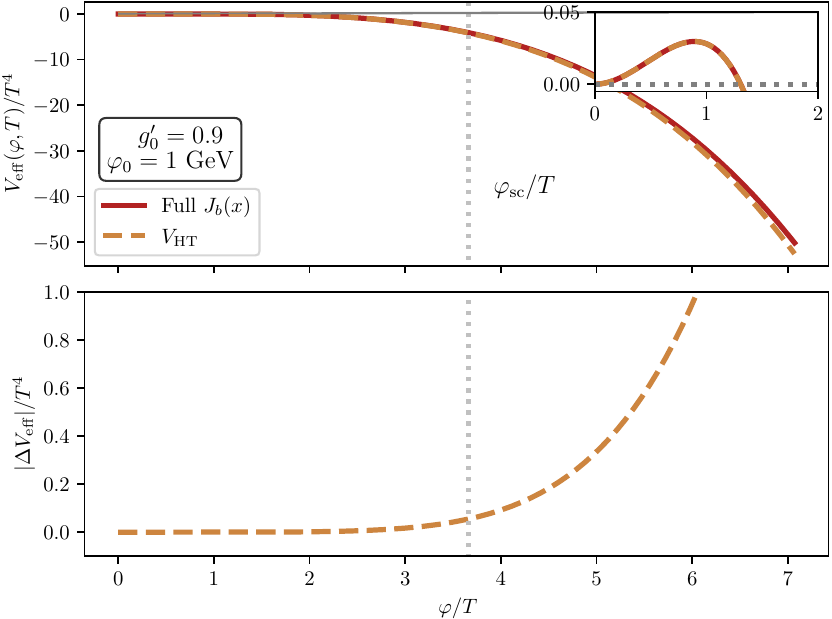}
    \caption{Effective potential without Daisy contributions using the full thermal function $J_b$ (red line) or its HT expansion from Eq.~(\ref{eq:HT_thermal_func}) (dashed yellow line), for two different values of the gauge coupling $g'_0$ at the reference scale $\mu_0=g'_0\varphi_0$. The lower part of the plots shows the relative error, which starts to be large for large values of $\varphi/T$ where the HT expansion is not justified. The vertical dotted silver line corresponds to the escape point of the tunneling process, $\varphi_\mathrm{sc}$ (see Section~\ref{sec:tunneling}). The temperature is set to $T=\varphi_0/100$, which only affects the position of $\varphi_{sc}/T$.}
    \label{fig:Comparison_HT}
\end{figure}
We show the comparison between the use of the full $J_b(x)$ function in $V_{\mathrm{eff}}$ (solid red line) and its HT approximation from Eq.~(\ref{eq:HT_thermal_func}) (dashed yellow line) in Fig.~\ref{fig:Comparison_HT}. At this stage we neglect $V_{\mathrm{Daisy}}$. We plot results for two reference values of the gauge coupling $g'_0$, set at the scale $\mu_0=m_{Z',0}$. We then compute $V_{\mathrm{eff}}$ at the scale $\mu^2 = \pi^2 T^2$, which is motivated by the comparison between the $3D$-NLO effective potential~\cite{Christiansen:2025xhv} and the $4D$ result we study. Accordingly, we run the relevant couplings from $\mu_0$ to $\mu$ by solving the RG equations discussed in Subsection~\ref{sec:RGE}. The temperature is fixed at $T=\varphi_0/100$. As can be seen in the upper panels, smaller gauge couplings tend to displace the barrier to larger values of $\varphi/T$. We can estimate the width of the barrier $\varphi_b$ by using the HT expansion~\cite{Salvio:2023blb}, finding
\begin{equation}
	\varphi_b \simeq - \frac{2 K(T)}{3\lambda(T)}\left[1-\sqrt{1+ \frac{9 m^2(T)\lambda(T)}{2K(T)^2}}\right]\,,
    \label{eq:width_barrier}
\end{equation}
which depends on the gauge coupling as $\varphi_b\sim 1/g'_0$. 

The vertical dotted gray line in Fig.~\ref{fig:Comparison_HT} corresponds to the field value at which tunneling takes place, $\varphi_{\mathrm{sc}}$, and from which it just ``rolls down'' to the true minimum. More details about the physical interpretation of $\varphi_{\mathrm{sc}}$ can be found in Section~\ref{sec:tunneling}. Beyond this point, the shape of the potential has no impact when computing the action $S_3$, quantity which drives the tunneling process. In the lower panels we show the absolute error committed by using the HT expansion, normalized by $T^4$, which clearly shows the excellent agreement between the full and approximated solutions until the escape point $\varphi_{\mathrm{sc}}$. Although the error grows quickly for large field values at which the HT expansion is no longer justified, we stress that the relative error is at most $\mathcal{O}(2\%)$ for the region of field space we are interested in, i.e. $\varphi\lesssim \varphi_{\mathrm{sc}}$. This is clearly shown in the lower-right panel, where the larger value of $g'_0$ drifts the potential barrier to smaller field values. Although there is poor agreement for $\varphi/T\gtrsim 5$, this region plays no role in the tunneling process. 

\subsection{Daisy contribution}\label{sec:Daisy}
It is fundamental to include Daisy resummation when considering the HT limit with boson species, as it cannot be neglected when studying the tunneling process~\cite{Kierkla:2023von}. For classically scale-invariant scenarios, we can parametrize the boson thermal mass as $\Pi_b(T)=C_b g'^2T^2$, which for the $U(1)^{\prime}$ scale-invariant model we study specifies to $C_{Z'}=C_b=1/3$ at leading order. In order to use (semi-) analytical solutions for the bounce action $S_3$, we wish to write the full effective potential as a polynomial on $\varphi$. This means that we also need to find an approximation for the following non-polynomial term
\begin{equation}
	(m_b^2(\varphi)+\Pi_b(T))^{3/2}=\Pi_b^{3/2}(T)\left(1+\frac{\varphi^2}{C_b T^2}\right)^{3/2}\equiv \Pi_b^{3/2}(T)f(x)\,,
	\label{eq:term_Daisy_expansion}
\end{equation}
where we have defined $x\equiv \varphi/T$ and $f(x)\equiv(1+x^2/C_b)^{3/2}$, such that we can write $V_{\mathrm{Daisy}}(\varphi,T)\sim \sum_{n=0}^4 c_n(T) \varphi^n$. Although extensively found in the literature, Taylor expanding $f(x)$ assuming $x\ll 1$ is only valid for $x\ll C_b$. In contrast to the HT expansion of the thermal functions we previously introduced, which is valid up to $x\sim 1/g'$, the Taylor expansion of the Daisy contribution would only be valid for $x< C_{Z'}=1/3$. As already shown in Fig.~\ref{fig:Comparison_HT}, we need to study the effective potential up to $x\simeq \mathcal{O}(5)$, which renders this Taylor expansion unreliable to describe the FOPT dynamics. 

Instead, we advocate here for the projection of the non-polynomial Daisy term in Eq.~(\ref{eq:term_Daisy_expansion}) into a basis of orthogonal polynomials. We discuss two different possibilities in the following:
\begin{itemize}
	\item Legendre polynomials:
	
	We can decompose $f(x)$ in terms of Legendre polynomials $P_n(x)$, as $f(x)\simeq f^{\mathrm{Leg}}(x)= \sum_{n=0}^4c_{n}^\mathrm{Leg}P_n(t(x))$.\footnote{The orthogonality condition for Legendre polynomials of order $n$, $P_n(x)$, is given by $\int_{-1}^1 dt ~ P_n(t)P_m(t) =\frac{2}{2n+1}\delta_{nm}$.} Note that we shift the polynomials from the interval $t\in [-1,1]$ in which they are defined onto $x\in[0,b]$, using the transformation $t(x)\equiv 2x/b-1$. The constant $b$ is chosen to be $b\sim\mathcal{O}(5)$ to roughly cover the relevant region for tunneling (see Fig.~\ref{fig:Comparison_HT}). The coefficients are then given by
	\begin{equation}
		c_n^{\mathrm{Leg}}=\frac{2n+1}{2}\int_{-1}^1P_n(t)f(x(t))dt\,.
	\end{equation}
	Note that with this procedure a tadpole term is introduced, which is nevertheless removed by a field redefinition we specify later, and also in Appendix~\ref{app:Daisy_poly_coefficients}.
    
	\item Gram-Schmidt decomposition:
	
	In this case we explicitly decompose $f(x)\simeq f^{\mathrm{GS}}(x)= 1 + \sum_{n=2}^4 c_n^{\mathrm{GS}} x^n$. In contrast to the previous case, this approach avoids the introduction of a tadpole, making its use straightforward. We can also find analytical expressions for the $c_n^{\mathrm{GS}}$ by minimizing the following quantity
	\begin{equation}
		d=\int_0^{b}dx ~ (f(x)-f^{\mathrm{GS}}(x))^2\,,
	\end{equation}
	with respect to the coefficients $c_n^{\mathrm{GS}}$. Once again, we choose $b\sim\mathcal{O}(5)$ to cover the relevant region for the computation of the bounce action. 
\end{itemize}

We write down the explicit expressions for both $c_n^{\mathrm{Leg}}$ and $c_n^{\mathrm{GS}}$ in Appendix~\ref{app:Daisy_poly_coefficients}, and leave in the following their numerical values taking $b=5$ for the $U(1)^{\prime}$ model with $C_{b}=C_{Z'}=1/3$:
\begin{equation}
    \begin{split}
        &c_0^{\mathrm{Leg}}=169.0\,,\;c_1^{\mathrm{Leg}}=298.6\,,\;c_2^{\mathrm{Leg}}=162.6\,,\;c_3^{\mathrm{Leg}}=32.32\,,\;c_4^{\mathrm{Leg}}=0.1066\,,\\
        &c_0^{\mathrm{GS}}=1.0\,,\;c_1^{\mathrm{GS}}=0.0\,,\;c_2^{\mathrm{GS}}=2.029\,,\;c_3^{\mathrm{GS}}=4.614\,,\;c_4^{\mathrm{GS}}=0.05503\,.        
    \end{split}
\end{equation}
Our results apply for $C_b>0$, which is our case study. Nevertheless, the same procedure can be followed to obtain the coefficients for $C_b<0$ (with the additional condition that $b^2+C_b<0$), although we do not write these results explicitly. For scalar contributions one would just need to change the particular value of $C_b$. 

Using any of the two approximations, we can finally write the Daisy contribution as a polynomial
\begin{equation}
    \begin{split}
        V_{\mathrm{Daisy}}^{\mathrm{Leg}}(\varphi,T) = & V_{\mathrm{Leg},0}+\nu^3_{\mathrm{Leg}}\varphi+\frac{1}{2}m^2_{\mathrm{Leg}}\varphi^2-\frac{1}{3}K_{\mathrm{Leg}}\varphi^3-\frac{1}{4}\lambda_{\mathrm{Leg}}\varphi^4\,,\\
        V_{\mathrm{Daisy}}^{\mathrm{GS}}(\varphi,T) = & V_{\mathrm{GS},0}+\frac{1}{2}m^2_{\mathrm{GS}}\varphi^2-\frac{1}{3}K_{\mathrm{GS}}\varphi^3-\frac{1}{4}\lambda_{\mathrm{GS}}\varphi^4\,.
    \end{split}
    \label{eq:Daisy_Poly}
\end{equation}
The polynomial parameters in Eq.~(\ref{eq:Daisy_Poly}), which depend on the temperature, are also written explicitly in terms of $c_n^{\mathrm{Leg}\,(\mathrm{GS})}$ in Appendix~\ref{app:Daisy_poly_coefficients}. Choosing the Gram-Schmidt approximation for $V_{\mathrm{Daisy}}$, we can write the full effective potential from Eq.~(\ref{eq:polynomial_shape_pot}) simply as
\begin{equation}
    V_{\mathrm{eff}}^{\mathrm{GS}}(\varphi,T)\simeq V_0+V_{\mathrm{GS},0}+\frac{1}{2}(m^2+m^2_{\mathrm{GS}})\varphi^2-\frac{1}{3}(K+K_{\mathrm{GS}})\varphi^3-\frac{\lambda+\lambda_{\mathrm{GS}}}{4}\varphi^4\,,
    \label{eq:Veff_GS_final}
\end{equation}
which can be directly used to compute $S_3$ semi-analytically using the results of Ref.~\cite{Levi:2022bzt}. 

\begin{figure}
    \centering
    \includegraphics[width=0.49\linewidth]{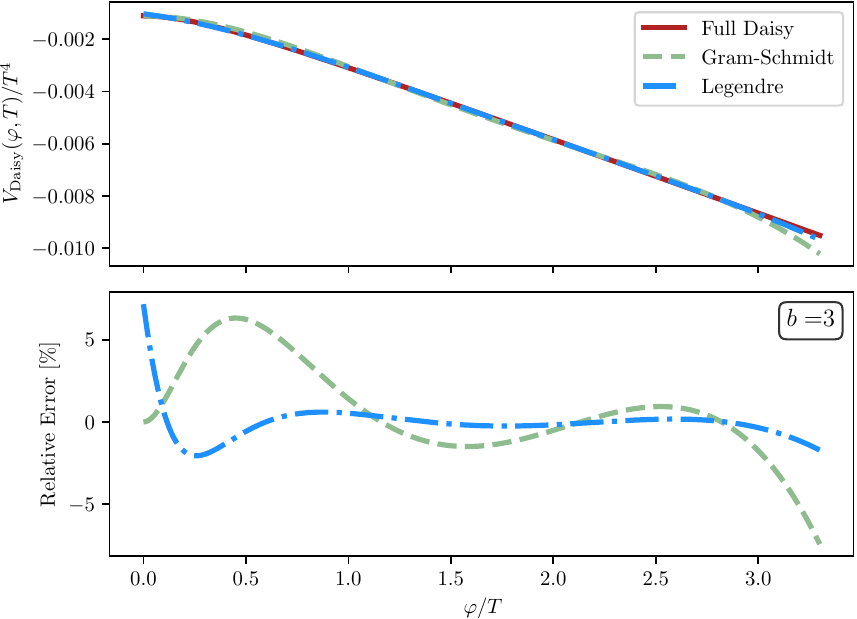}
    \includegraphics[width=0.49\linewidth]{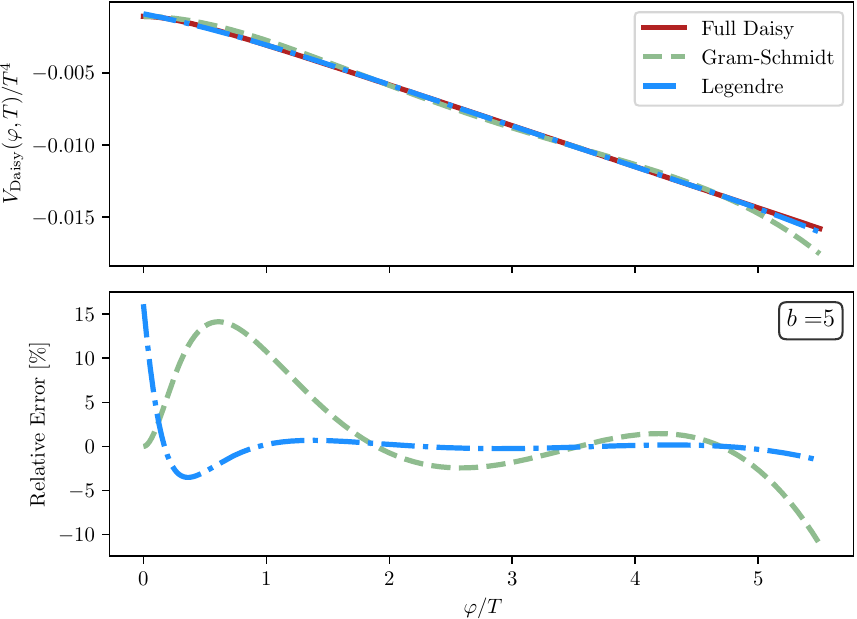}
    \caption{Comparison between the full Daisy contribution to the effective potential for the $U(1)^{\prime}$ model, shown as a solid red line, and the polynomial approximations. The lower panels show the relative error for each approximation, for $b=3$ (left) and $b=5$ (right). The dashed green line corresponds to the Gram-Schmidt procedure, while the dash-dotted blue line to the approximation in terms of Legendre polynomials.}
    \label{fig:comparison_Daisy_terms}
\end{figure}

In contrast, by using $V_{\mathrm{Daisy}}^{\mathrm{Leg}}$ we introduce a tadpole term that needs to be removed. This can be achieved by an appropriate field redefinition, $\varphi\rightarrow \tilde{\varphi}-w$, such that the effective potential in Eq.~(\ref{eq:polynomial_shape_pot}), written in terms of $\tilde{\varphi}$, does not present a linear term. The shift $w$ is implicitly given by the solution to the following cubic equation, whose explicit solutions are given in Appendix~\ref{app:Daisy_poly_coefficients}:
\begin{equation}
    \nu^3_{\mathrm{Leg}}-(m^2+m^2_{\mathrm{Leg}})w-(K+K_{\mathrm{Leg}})w^2+(\lambda+\lambda_{\mathrm{Leg}})w^3=0\,.
    \label{eq:implicit_cubic_tadpole}
\end{equation}
We choose the smallest field shift that removes the tadpole, such that the full effective potential is then
\begin{equation}
    \tilde{V}_{\mathrm{eff}}^{\mathrm{Leg}}(\tilde{\varphi},T)=\tilde{V}_{\mathrm{Leg},0}+\frac{1}{2}\tilde{m}^2_{\mathrm{Leg}}\tilde{\varphi}^2-\frac{1}{3}\tilde{K}_{\mathrm{Leg}}\tilde{\varphi}^3-\frac{1}{4}\tilde{\lambda}_{\mathrm{Leg}}\tilde{\varphi}^4\,,
    \label{eq:Veff_Leg_final}
\end{equation}
with the parameters given by
\begin{equation}
    \begin{split}
        \tilde{V}_{\mathrm{Leg},0}\equiv &V_0+V_{\mathrm{Leg},0}-\nu^3_{\mathrm{Leg}} w + \frac{1}{2}(m^2+m_{\mathrm{Leg}}^2)w^2+\frac{K+K_{\mathrm{Leg}}}{3}w^3-\frac{\lambda+\lambda_{\mathrm{Leg}}}{4}w^4\,,\\
        \tilde{m}^2_{\mathrm{Leg}}\equiv &(m^2+m_{\mathrm{Leg}}^2)+2(K+K_{\mathrm{Leg}})w-3(\lambda+\lambda_{\mathrm{Leg}})w^2\,,\\
        \tilde{K}_{\mathrm{Leg}}\equiv &(K+K_{\mathrm{Leg}}) - 3 (\lambda+\lambda_{\mathrm{Leg}})w\,,\\
        \tilde{\lambda}_{\mathrm{Leg}}\equiv & (\lambda+\lambda_{\mathrm{Leg}})\,.
    \end{split}
\end{equation}
The potential in Eq.~(\ref{eq:Veff_Leg_final}) can now also be used to compute the bounce action semi-analytically. The results, which depend on the size of the interval for $\varphi/T$ over which we perform the projection, are shown in Fig.~\ref{fig:comparison_Daisy_terms} for the $U(1)^{\prime}$ model with $C_{Z'}=1/3$ and two choices of $b$: we take $b=3$ in the left panel, and $b=5$ in the right one. As expected, although the qualitative behavior is similar in both cases, the further away from the origin we try to describe the Daisy contribution in terms of polynomials, the larger the error we make close to the false vacuum. Moreover, although overall $V_{\mathrm{Daisy}}^{\mathrm{Leg}}$ (dash-dotted blue line) describes $V_{\mathrm{Daisy}}$ (solid red line) better than $V_{\mathrm{Daisy}}^{\mathrm{GS}}$ (dashed green line), the latter instead exactly reproduces the Daisy contribution close to the false vacuum. This is because the basis of Legendre polynomials always has $\varphi$-independent terms that make it impossible to exactly reproduce $V_{\mathrm{Daisy}}(\varphi\rightarrow0,T)$. In contrast, $V_{\mathrm{Daisy}}^{\mathrm{GS}}$ is explicitly built to give the same result as the full Daisy contribution at the origin. 

\subsection{Renormalization scale}\label{sec:RGE}
In our previous discussion, we highlighted the dependence of $V_{\mathrm{eff}}(\varphi,T)$ on the choice of the renormalization scale $\mu^2$. As advocated in Ref.~\cite{Kierkla:2023von}, and as justified when studying the HT expansion, we set the renormalization scale to
\begin{equation}
    \mu = \mathrm{max}\left[m_{Z'}(\varphi),\,\sqrt{\kappa} T\right]\,,
    \label{eq:choice_mu}
\end{equation}
where $\kappa$ is just a proportionality constant, which we set to $\kappa = \pi^2$. This value has recently been found to best reproduce the behavior of the effective potential in the dimensionally reduced theory at NLO~\cite{Christiansen:2025xhv}, which we take as a reference given its milder renormalization scale dependence and its use in state-of-the-art computations. Note that this is just a choice at this stage, and that other options could also be well motivated, such as $\kappa = a_b$ which completely removes the logarithm in the HT expansion in Eq.~(\ref{eq:polynomial_params_NoDaisy}). 

The choice of $\mu$ in Eq.~(\ref{eq:choice_mu}) means that for field values near the origin and the potential barrier, for which $\varphi\lesssim T$, we change the renormalization scale according to the temperature. Instead, for field values near the true minimum $\varphi_0$, and focusing for the moment on supercooled FOPT for which we expect $T\ll\varphi_0$ for the milestone temperatures, the appropriate choice for the renormalization scale corresponds to the $Z'$ mass. This distinction naturally sets two scales at which we need to study the effective potential in order to describe FOPT dynamics~\cite{Kierkla:2023von}. It is therefore fundamental to study the full RGE-improved effective potential~\cite{Meissner:2008uw,Chataignier:2018kay,Chataignier:2018aud,Ellis:2020nnr}, which is equivalent to studying the $3D$ effective field theory at leading order~\cite{Lewicki:2024xan}. While the mitigation of the renormalization scale dependence of $V_{\mathrm{eff}}(\varphi,T)$ is an important issue when precisely studying GW predictions~\cite{Gould:2021oba,Athron:2023xlk,Athron:2023rfq,Croon:2020cgk,Lewicki:2024xan}, it goes beyond the scope of our work. We highlight that the approximations we have introduced so far, as well as the ones we will perform to compute the fraction of the Universe in the false vacuum, are independent of the method used to handle the renormalization scale dependence and stress that these dependencies can be introduced in our framework by adjusting the polynomial coefficients for the effective potential. Whichever method is adopted, its corresponding uncertainty from renormalization scale dependence will always be present. 

Next, we discuss in detail the RGE of the relevant couplings for the $U(1)^{\prime}$ model that we use as an example. Once we set the $Z'$ mass and the $U(1)^{\prime}$ gauge coupling at the reference scale $\mu_0=m_{Z',0}= g'_0 \varphi_0$, it is easy to check that the tree-level quartic coupling $\lambda$ (see Eq.~(\ref{eq:tree_level_pot})) needs to satisfy
\begin{equation}
    \lambda_0=\frac{g'^4_0}{16\pi^2}\,,
    \label{eq:relation_lambda_gprime}
\end{equation}
so that we have radiative symmetry breaking (RSB), and thus a minimum at $\varphi_0$. In order to compute the value of the couplings at different renormalization scales, we need to solve their RGE-flows, given by
\begin{equation}
    \begin{cases}
        \frac{dg'}{dt}=&\frac{1}{16\pi^2}\frac{g'^3}{3}\,,\\
        \frac{d\lambda}{dt}=&\frac{1}{8\pi^2}\left(3g'^4-6g'^2\lambda+10\lambda^2\right)\,,
    \end{cases}
    \label{eq:RGE_running}
\end{equation}
where $t\equiv \log{\mu/\mu_0}$ and we left the $t$-dependence of the couplings implicit. We have computed the $\beta$ functions (and anomalous dimension) using \texttt{PyR@TE}~\cite{Poole:2019kcm} and checked them against Ref.~\cite{Goncalves:2025uwh}.
The exact analytic solutions to Eq.~(\ref{eq:RGE_running}) can be written as
\begin{equation}
    \begin{split}
        g'^2(t)=&\frac{24\pi^2g'^2_0}{24\pi^2-g'^2_0t}\,,\\
        \lambda(t) =& \frac{g'^2(t)}{60} \left[19 + \sqrt{719} \tan \left ( \tan^{-1}{\left (
\dfrac{120 \dfrac{\lambda_0}{g'^2_0} - 38}{2\sqrt{719}} \right )} + \dfrac{\sqrt{719}}{2} \log{\left( \dfrac{g'^2(t)}{g'^2_0} \right )} \right )\right]\,.\\
    \end{split}
    \label{eq:explicit_RGE_exact}
\end{equation}
Finally, the field renormalization is given by $\varphi\rightarrow Z^{1/2}_{\varphi}(t) \varphi'$, with
\begin{equation}
    Z_{\varphi}(t) = \exp{\left[-\frac{1}{2}\int_0^t dx ~\gamma(x)\right]}\,,
    \label{eq:field_renom}
\end{equation}
where $\gamma(x)=-3g'^2(x)/(16\pi^2)$ is the anomalous dimension. We can once again perform the integration analytically and find
\begin{equation}
    Z_{\varphi}(t)=\exp\left[-\frac{9}{4}\log{\left(\frac{g'^2_0}{g'^2(t)}\right)}\right]\,.
    \label{eq:field_renormalization_solution}
\end{equation}

While we explicitly include the full running of $\lambda(t)$ and $g'(t)$ in our computations, we neglect the correction arising from the field renormalization given that we find $Z_{\varphi}(t)\sim \mathcal{O}(1)$ at percent precision. However, we highlight that its full inclusion is straightforward and would not prevent us from writing $V_{\mathrm{eff}}(\varphi',T)$ as a polynomial, and thus from exploiting the analytic results on $S_3$. Including the field renormalization would be tantamount to making the following replacements in the polynomial coefficients:
\begin{equation}
    \begin{split}
        m^2(T)\rightarrow m^2(T)/Z_{\varphi}(t)\,,&\quad K(T)\rightarrow K(T)/Z_{\varphi}(t)\,,\\
        \lambda(T)\rightarrow \lambda(T)/Z_{\varphi}(t)\,,&\quad S_3(T)\rightarrow S_3(T)Z^{3/2}_{\varphi}(t)\,,
    \end{split}
    \label{eq:replacements_field_renom}
\end{equation}
where the coefficients' dependence on $t$ is left implicit. The correction to the action arises because $S_3$ is computed assuming that the kinetic term for the scalar is canonically normalized, and therefore we need to shift its value.

\section{Tunneling action}\label{sec:tunneling}
In this section, we summarize previous semi-analytical results from Ref.~\cite{Levi:2022bzt} in which the bounce action for a polynomial potential was analyzed. Neglecting the tunneling rate in vacuum, which is negligible for the regions of parameter space we are interested in, namely those with $g'\lesssim 1$~\cite{Levi:2022bzt}, the probability  to transition per unit volume in the early Universe, from a metastable minimum $\varphi_+$ to a deeper one $\varphi_0$ with $V_{\mathrm{eff}}(\varphi_+)>V_{\mathrm{eff}}(\varphi_0)$, is given by~\cite{Gould:2021ccf,Hirvonen:2024rfg}~\footnote{See Ref.~\cite{Kierkla:2025qyz} for a state-of-the-art analysis of the nucleation rate including the explicit computation of the fluctuation determinant and its impact on the subsequent GW predictions.}
\begin{equation}
    \frac{\Gamma}{V}= T^4\left(\frac{S_3(T)}{2\pi T}\right)^{3/2}e^{-S_3(T)/T}\,.
    \label{eq:tunneling_probability}
\end{equation}
In order to find the bounce action $S_3(T)$, we need to solve the equations of motion for the scalar~\cite{Coleman:1977py,Linde:1980tt}\footnote{An equivalent formalism can be used in order to describe the tunneling process and which can be reduced to a minimization procedure instead of solving an ODE~\cite{Espinosa:2018hue,Espinosa:2018szu}, being numerically more stable.}
\begin{equation}
    \frac{d^2\varphi}{d\rho^2}+\frac{D-1}{\rho}\frac{d\varphi}{d\rho} = \frac{dV_{\mathrm{eff}}}{d\varphi}\,,
    \label{eq:bounce_eq}
\end{equation}
with the appropriate boundary conditions
\begin{equation}
    \begin{split}
        \frac{d\varphi}{d\rho}\Bigg|_{\rho=0}=0\,,\quad \lim_{\rho\rightarrow \infty}\varphi=\varphi_+\,,
    \end{split}
    \label{eq:bounce_boundary}
\end{equation}
where $\rho\equiv|\vec{x}|$. We are interested in the transition at finite temperature, such that $D=3$. In full generality, one needs to solve Eq.~(\ref{eq:bounce_eq}) numerically, using for example a shooting method which amounts to finding the initial field configuration $\varphi(\rho=0) = \varphi_{\mathrm{sc}}$ for which the boundary conditions in Eq.~(\ref{eq:bounce_boundary}) are met. Once the solution $\varphi(\rho)$ is found, the three-dimensional bounce action is given by
\begin{equation}
    S_3(T) = 4 \pi\int_0^{\infty}d\rho ~\rho^2\left[\frac{1}{2}\left(\frac{d\varphi}{d\rho}\right)^2+V_{\mathrm{eff}}(\varphi,T)\right]\,.
\end{equation}

Applying the results from Section~\ref{sec:Veff_poly}, allowing us to write the effective potential in terms of a quartic polynomial as $V_{\mathrm{eff}}(\varphi,T) = V_0(T) + m^2(T)\varphi^2/2-K(T)\varphi^3/3-\lambda(T)\varphi^4/4$, we can now use the results of Ref.~\cite{Levi:2022bzt}, where a fit to $S_3(T)$ in terms of $\kappa\equiv - \lambda(T) m^2(T) / K(T)^2$ was found. The following result was obtained
\begin{equation}
    S_3(T)\simeq 
    \begin{cases}
        \frac{32 \pi}{729} \frac{m^3}{K^2 (\kappa - 2/9)^2} B_3^+(\kappa)\,, & \kappa > 0 \\ 
        \frac{6 \pi m}{\lambda}B_3^-(\kappa)\,, & \kappa \leq 0
    \end{cases}\,,
    \label{eq:fit_action}
\end{equation}
where the functions $B_3^{\pm}(\kappa)$ can be found in Ref.~\cite{Levi:2022bzt}. Thus, profiting from the approximations discussed in Section~\ref{sec:Veff_poly} for the effective potential, one can compute $S_3(T)$ without solving Eq.~(\ref{eq:bounce_eq}) numerically for each point in parameter space. Note that $\kappa=2/9$ corresponds to the case of exactly degenerate vacua, for which analytical solutions are known in the thin-wall limit~\cite{Coleman:1977py}.

\begin{figure}
    \includegraphics[width=0.49\linewidth]{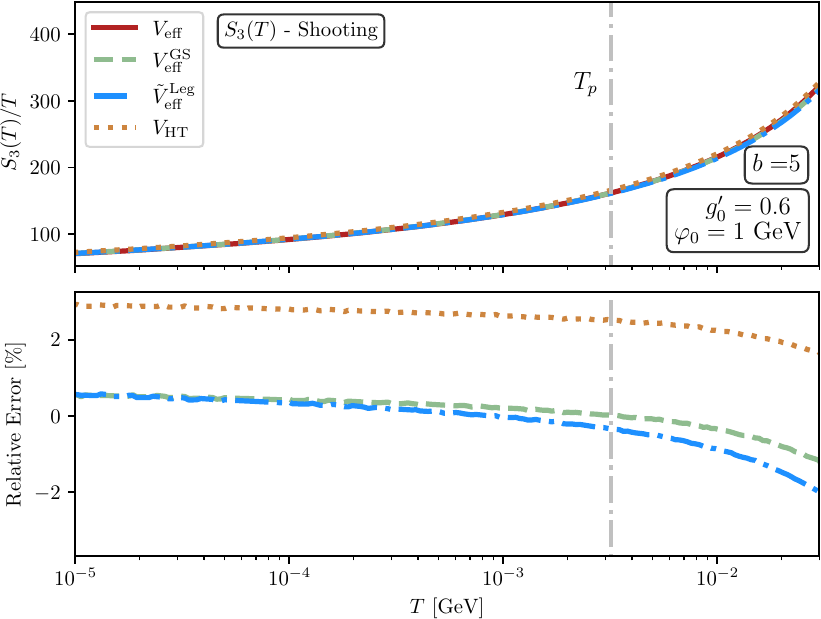}
    \includegraphics[width=0.49\linewidth]{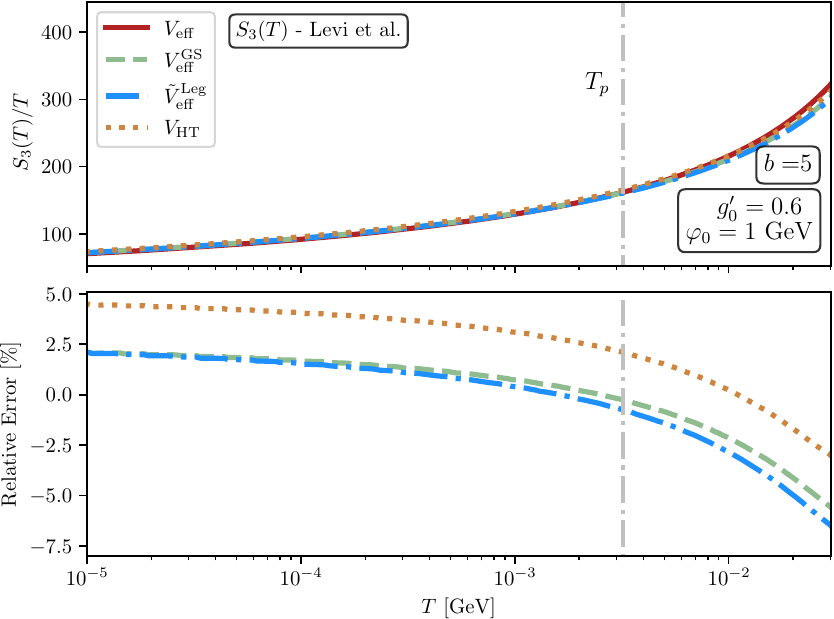}
    \caption{Bounce action for the $U(1)^{\prime}$ scale-invariant scenario using different approximations for the effective potential and the computation of $S_3(T)$. In both panels, the red solid line corresponds to the result using the full effective potential and computing the bounce action numerically with \texttt{CosmoTransitions}. The green dashed line corresponds to the effective potential in Eq.~(\ref{eq:Veff_GS_final}), the dash-dotted blue one to $V_{\mathrm{eff}}$ in Eq.~(\ref{eq:Veff_Leg_final}), and the dotted yellow line neglects the Daisy contribution. In the left panel we compare the impact of different approximations to $V_{\mathrm{eff}}$ on $S_3$, computed using the shooting method. In the right panel, we compare the values of $S_3$, obtained with the fit from Ref.~\cite{Levi:2022bzt}, for the polynomial potentials against the full result.} 
    \label{fig:comparison_actions}
\end{figure}
In the left panel of Fig.~\ref{fig:comparison_actions}, we compare how the different polynomial approximations for $V_{\mathrm{eff}}$ impact the value of $S_3(T)$. We compute the action with the extensively tested shooting routine from \texttt{CosmoTransitions}~\cite{Wainwright:2011kj} in every case. In the right panel, we instead compare the results for $S_3(T)$ from Eq.~(\ref{eq:fit_action}) using the polynomial approximations for $V_{\mathrm{eff}}$ against the full numerical computation without approximations. The full numerical result is shown as a solid red line, the result using the HT expansion together with the Gram-Schmidt (Legendre) approach for Daisy terms as a dashed green (dash-dotted blue) line, and the result neglecting altogether the Daisy contribution is presented as a dotted yellow line.

The results shown in the left panel allow us to assess the quality of each polynomial approximation in reproducing the full effective potential around the barrier, and consequently the tunneling action. As we already argued, the tunneling process is mostly sensitive to small field values for supercooled FOPTs, naturally falling in the HT regime. Therefore, it is critical to include Daisy resummation. This can be realized by comparing the results that describe the Daisy term as a polynomial with respect to the case in which we completely neglect the Daisy contribution. While $V_{\mathrm{eff}}^{\mathrm{GS}}$ and $\tilde{V}_{\mathrm{eff}}^{\mathrm{Leg}}$ tend to underestimate the action, introducing a relative error of at most $2\%$, we note that neglecting $V_{\mathrm{Daisy}}$ roughly overestimates $S_3$ by, at least, $2\%$. Note that $S_3$ enters exponentially in Eq.~\eqref{eq:tunneling_probability}, so that even a small relative error in $S_3$ can result in a much bigger one in $\Gamma/V$. Moreover, we can see how neglecting $V_{\mathrm{Daisy}}$ becomes a worse approximation for smaller temperatures, which correspond to larger supercooling linked to better observational prospects for GW detection. This is in contrast to our polynomial approximations, which actually perform better in this region. 

In the right panel we find a similar trend as in the left one. Neglecting Daisy terms worsens the quality of the results for smaller temperatures, while our polynomial approximations are better in that region. However, we find that the analytical action in Eq.~(\ref{eq:fit_action}), taken from Ref.~\cite{Levi:2022bzt}, tends to underestimate the value of $S_3$. Given that both $V_{\mathrm{eff}}^{\mathrm{GS}}$ and $\tilde{V}_{\mathrm{eff}}^{\mathrm{Leg}}$ also underestimate to some extent the value of $S_3$ (see left panel), the combination of both approximations worsens the final result down to around $\mathcal{O}(5\%)$. Nevertheless, comparing the left and right panels showing the relative error, it is obvious that, for the region $T\lesssim 10^{-2}\varphi_0$, our polynomial approximations to the effective potential introduce a negligible error compared to the one arising from the adoption of $S_3$ in Eq.~(\ref{eq:fit_action}). Instead, given that, neglecting the Daisy terms makes the potential barrier larger, thus overestimating the action, the effect of using $S_3$ for a polynomial potential in this case gets compensated and translates into a reduction of the overall error for some temperatures. Nevertheless, for large supercooling the error grows when neglecting Daisy contributions, regardless of whether we compute $S_3$ numerically or using instead Eq.~(\ref{eq:fit_action}).

\section{Milestone temperatures}\label{sec:temperatures}
We devote this section to the computation of the relevant temperatures for the study of the FOPT, namely the critical, nucleation, and percolation temperatures. When studying the percolation temperature $T_p$, which corresponds to the temperature at which GWs are produced, we will introduce an approximation that will allow us to compute the fraction of the Universe in the false vacuum analytically, greatly simplifying the study of FOPTs for phenomenological purposes. We highlight that, while we will study how well our polynomial approximation can describe the FOPT and compute the milestone temperatures using it, the following discussion and approximations are completely independent of the shape of $V_{\mathrm{eff}}$. They can be adopted in full generality once the tunneling rate, $\Gamma/V$ in Eq.~(\ref{eq:tunneling_probability}), is computed.

\subsection{Critical temperature}
At the critical temperature $T_c$, the effective potential develops two degenerate vacua separated by a potential barrier. Note that the position of both vacua depends on the temperature. The critical temperature determines the onset of the FOPT, and can be easily obtained by directly studying the effective potential. At this stage in the history of the FOPT, tunneling is not possible, and therefore $T_c$ only provides information about the highest temperature at which the FOPT could take place, but has little interest for the subsequent GW production. 

\subsection{Nucleation temperature}
The nucleation temperature is that at which approximately one bubble per Hubble volume has been nucleated. It can approximately be found studying the condition
\begin{equation}
    \Gamma(T_n) \simeq H(T_n)\,,
    \label{eq:definition_Tn}
\end{equation}
with the Hubble expansion rate during the FOPT given by
\begin{equation}
    H(T)^2 = \frac{1}{3 m_{\mathrm{Pl}}^2}\left(\frac{g_*\pi^2T^4}{30}+\Delta V_{\mathrm{eff}}\right)\simeq \frac{\rho_\mathrm{rad}(T)}{3m_{\mathrm{Pl}}^2}\left(1+\alpha\right)\,.
    \label{eq:Hubble}
\end{equation}
The first term in the parenthesis in Eq.~(\ref{eq:Hubble}) corresponds to the radiation energy density, with $g_*$ taking into account both the SM and the new $U(1)^\prime$ relativistic degrees of freedom, while the second represents the vacuum energy density with $\Delta V_{\mathrm{eff}} \equiv  V_{\mathrm{eff}}(0, T) - V_{\mathrm{eff}}(\varphi_0(T),T)$, where we assume that at high temperatures we have symmetry restoration and therefore one of the minima is at the origin. The second equality follows in the case of supercooled FOPTs, such that $\partial \Delta V/\partial T\ll \Delta V/T$ at the relevant temperatures for GW production. Finally, $m_\mathrm{Pl}=2.4\times 10^{18}$~GeV is the reduced Planck mass. 
While $T_n$ is broadly used as a proxy to study GW production due to its simplicity, numerical simulations show instead that GWs are actually produced at the percolation temperature~\cite{Athron:2023xlk} (see Subsection~\ref{sec:percolation_temp}). Moreover, there are examples in which we do not find a nucleated bubble per Hubble volume, i.e. there is no nucleation temperature, but still GWs are produced~\cite{Athron:2023mer}. 

\subsection{Percolation temperature}\label{sec:percolation_temp}
The percolation temperature corresponds to the moment in the evolution of the FOPT when 29\% of the Universe is in the true vacuum, at which GWs are produced~\cite{Athron:2023xlk,Athron:2023mer}. We thus need to study now the fraction of the Universe in the false vacuum $P_f(T)=e^{-I(T)}$, with the exponent given by~\cite{Athron:2023xlk}
\begin{equation}
    I(T)\equiv \frac{4\pi v_w^3}{3}\int_{T}^{T_c}dT'\frac{\Gamma (T')}{T'^4H(T')}\left(\int_{T}^{T'}\frac{dT''}{H(T'')}\right)^3\,.
    \label{eq:percolation_full}
\end{equation}
Note that Eq.~(\ref{eq:percolation_full}) implicitly assumes that $dT/dt = - TH(T)$, and that the velocity at which the nucleated bubbles expand, $v_w$, is constant. 

The percolation temperature $T_p$ is then given by the condition $I(T_p)=  -\log{0.71}\simeq 0.34$. While obtaining the nucleation temperature only requires to find the action, something which can already be numerically expensive, computing the percolation temperature instead entails the double integral in Eq.~(\ref{eq:percolation_full}) with the unknown lower integration limits for which the condition $I(T_p)=0.34$ is satisfied. For fast FOPT we expect $T_n\sim T_p$, such that we can typically use the nucleation temperature. However, this does not need to be the case for a supercooled FOPT, and therefore computing the percolation temperature becomes critical~\cite{Athron:2023xlk}. 

In the following, we will introduce some approximations that allow us to determine $I(T)$ analytically via a Laplace approximation, reducing the computation of the percolation temperature from a double integration with unknown limits to a root-finding problem, similar to what is done for the nucleation temperature in Eq.~(\ref{eq:definition_Tn}). Overall, we are interested in integrating Eq.~(\ref{eq:percolation_full}) analytically, for which we first need to perform the inner integration. In order to ease the subsequent discussion, we define
\begin{equation}
    f(T,T')\equiv \frac{4\pi v_w^3}{3}\frac{1}{H(T')}\left(\int_{T}^{T'}\frac{dT''}{H(T'')}\right)^3\,.
\end{equation}
Depending on the energy density dominating the expansion rate during the FOPT, the shape of $f(T,T')$ will differ. We consider the case of fast transitions, for which we expect radiation domination, supercooled FOPTs with vacuum domination, and the intermediate case where both contributions are taken into account.
\begin{itemize}
    \item \textbf{Radiation domination:}
    
    If the free energy related to the effective potential is not very large, such that $\alpha \ll 1$, the transition takes place during radiation domination. In this case the Hubble expansion rate is given by $H(T)\simeq g_*^{1/2}\pi T^2/(3\sqrt{10} m_\mathrm{Pl})$,
    such that we find
    \begin{equation}
        f_{\mathrm{rad}}(T,T')=\frac{4\pi v_w^3}{3}\frac{8100m_\mathrm{Pl}^4}{g_*^2\pi^4}\frac{1}{T'^2}\left(\frac{1}{T}-\frac{1}{T'}\right)^3\,,
    \label{eq:fp_Radiation}
    \end{equation}
    which has a maximum at $T'=5T/2$. Note that we assumed that $g_*$ is constant, which should be a good approximation for fast transitions. Nevertheless, this result can be extended to include the temperature dependence of $g_*(T)$ by splitting the integration interval over relevant temperature regions in which $g_*$ is approximately constant.

    \item \textbf{Vacuum domination:}

    During vacuum domination, which will be the case whenever $\alpha\gg 1$, we can approximate $H(T)\simeq \Delta V_{\mathrm{eff}}^{1/2}/(\sqrt{3}m_\mathrm{Pl})$. As previously argued, in the presence of supercooling the temperature dependence of the effective potential is not relevant when computing the energy difference between vacua, which makes the integration of the Hubble expansion rate trivial using the $T=0$ result. Using this approximation we can identify in this case
    \begin{equation}
        f_{\mathrm{vac}}(T,T')=\frac{4\pi v_w^3}{3}\frac{9m_\mathrm{Pl}^4}{\Delta V_{\mathrm{eff}}^2}(T'-T)^3\,.
        \label{eq:fp_Vacuum}
    \end{equation}
    
    \item \textbf{General case:}

    Neglecting the matter contribution, we write the Hubble expansion rate as $H(T)=\sqrt{A+B T^4}$ by identifying $A\equiv \Delta V_{\mathrm{eff}}/(3m_{\mathrm{Pl}}^2)$ and $B\equiv g_*\pi^2/(90m_{\mathrm{Pl}}^2)$. Even in this case it is still possible to perform the inner integral in Eq.~(\ref{eq:percolation_full}) analytically. We find
    \begin{equation}
        \begin{split}
            g(T,T')\equiv& \int_{T}^{T'}\frac{dT''}{H(T'')}=
        \frac{-i}{(-A B)^{1/4}}\Bigg[F\left(i  \sinh^{-1}{\left(\left(-\frac{B}{A}\right)^{1/4}T'\right)}\Big|-1\right)\\
        &-F\left(i  \sinh^{-1}{\left(\left(-\frac{B}{A}\right)^{1/4}T\right)}\Big|-1\right)\Bigg]\,,
        \end{split}
        \label{eq:int_H_full}
    \end{equation}
    where $F(x|m)$ is the incomplete elliptic integral of the first kind. Using the result in Eq.~(\ref{eq:int_H_full}), we arrive at
    \begin{equation}
        f_{\mathrm{full}}(T,T')=\frac{4\pi v_w^3}{3}\frac{g(T,T')^3}{H(T')}\,.
        \label{eq:fp_Full}
    \end{equation}

\end{itemize}

\begin{figure}
    \centering
    \includegraphics[width=0.49\linewidth]{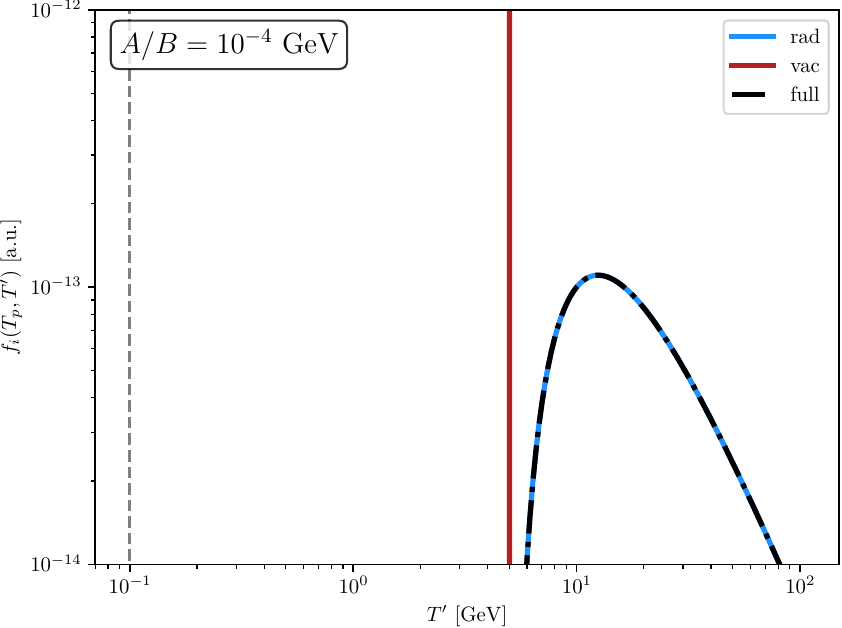}
    \includegraphics[width=0.49\linewidth]{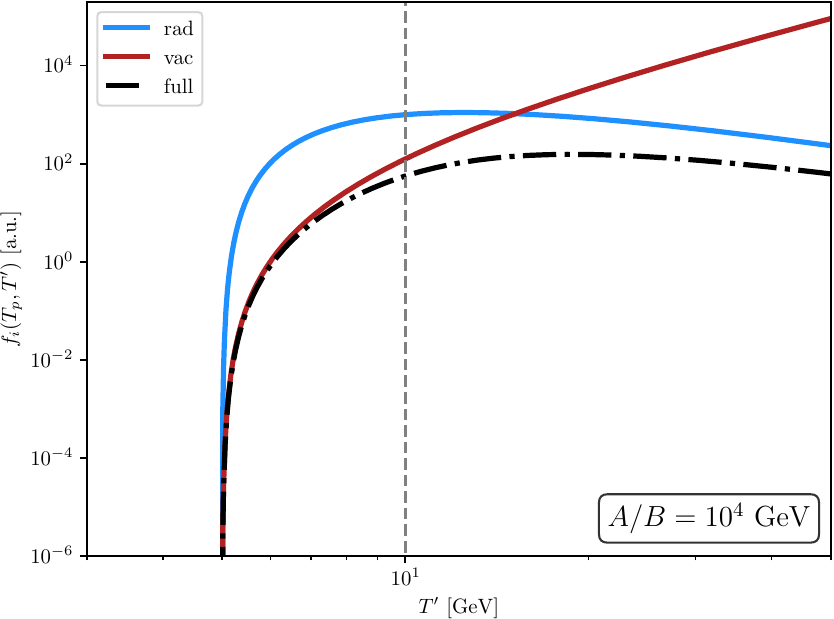}
    \caption{Comparison of the result for $f_i(T,T')$ assuming radiation domination (blue), vacuum domination (red) or using the full Hubble expansion rate (black dash-dotted). The vertical dashed gray line represents the temperature below which vacuum energy dominates the expansion of the Universe, given by $(A/B)^{1/4}$ with $A\,(B)$ defined above Eq.~(\ref{eq:int_H_full}). In the left panel radiation dominates the energy density of the Universe, while in the right one we have an intermediate case.}
    \label{fig:cases_f_i}
\end{figure}

We show in Fig.~\ref{fig:cases_f_i} the shape for the function $f_i(T,T')$ with $i=$rad, vac, and full taking into account only radiation (blue), only vacuum energy (red), and both (dash-dotted black), respectively. The dashed vertical gray line represents the temperature at which vacuum energy starts to dominate the energy density of the Universe, and therefore its expansion. The left panel corresponds to a case with strong radiation domination, while the right shows an example where the contribution from $\Delta V_{\mathrm{eff}}$ slightly dominates the energy budget of the Universe. Even though the full result interpolates very well between the cases of vacuum domination and radiation domination around the equality temperature in the right panel, we find in general that it is a good approximation to consider either radiation domination for fast FOPTs, or vacuum domination for the computation of $T_p$ in supercooled ones. 

Once the shape of the function $f(T, T')$ is known, we can turn to solve the final integration in Eq.~(\ref{eq:percolation_full}), which we write as
\begin{equation}
    I(T)=\int_{T}^{T_c}dT'f(T,T')\frac{\Gamma(T')}{T'^4}=\int_{T}^{T_c}dT'e^{-G(T,T')}\,,
    \label{eq:I_Tp_gauss_1}
\end{equation}
such that we can perform the integration independently of the particular form of $f(T,T')$, and where we have introduced, using Eq.~(\ref{eq:tunneling_probability}), the function $G(T,T')$ given by
\begin{equation}
    G(T,T')\equiv \frac{S_3(T')}{T'}-\log{f(T,T')}-\frac{3}{2}\log{\frac{S_3(T')}{2\pi T'}}\,.
    \label{eq:function_exp_percolation}
\end{equation}

The key point to take into account in the following is to notice that the function $f(T,T')$ identically vanishes when $T'\rightarrow T$. For $T'>T$, we find that $f(T,T')$ either has a maximum and again vanishes as $T'\rightarrow \infty$, or that it grows, at most, as a polynomial.\footnote{The result using $f_{\mathrm{full}}(T,T')$ is less straightforward but should fall between the limiting cases of vacuum and radiation domination.} These properties, together with the fact that $\Gamma(T')/T'^4$ is either monotonically decreasing, or again has a maximum for $T'<T_c$, translate into the existence of a maximum of the integrand in Eq.~(\ref{eq:I_Tp_gauss_1}) at some temperature $T_*\in (T,T_c)$. Note that $T_*$ would instead be a minimum of $G(T,T')$. We can then approximate the integral by Taylor expanding $G(T,T')$ to second order around its minimum at $T'=T_*$, arriving at the following expression for $I(T_p)$:
\begin{equation}
    \begin{split}
        I(T_p)\simeq &e^{-G_*}\sqrt{\frac{\pi}{2G_*^{''}}}
        \left[\mathrm{erf}\left(\sqrt{\frac{G_*^{''}}{2}}(T_c-T_*)\right)-\mathrm{erf}\left(\sqrt{\frac{G_*^{''}}{2}}(T_p-T_*)\right)\right]\,,
    \end{split}
    \label{eq:I_Tp_full}
\end{equation}
where primed quantities denote partial derivatives with respect to $T'$, such that $G_*\equiv G(T_p,T_*)$ and $G_*^{''}\equiv \partial^2G(T_p,T')/\partial T'^2|_{T_*}$. We find that the term proportional to $\log{S_3(T')/T'}$ in Eq.~(\ref{eq:function_exp_percolation}) does not change the results and could be directly evaluated at $T'=T_*$. Thanks to this approximation, we can now find the percolation temperature just by finding the root to the equation $I(T_p)=0.34$. Eq.~(\ref{eq:I_Tp_full}) can be further simplified by extending the integration limits to $T_c-T_*\rightarrow \infty$ and $T_p-T_*\rightarrow -\infty$. While of course this is not physically possible, it greatly simplifies the expression for $I(T_p)$, only introducing additional contributions that are nevertheless exponentially suppressed and thus negligible. We indeed find that the difference in the resulting percolation temperature using Eq.~(\ref{eq:I_Tp_full}) or the expressions in Eq.~(\ref{eq:approx_percolation_final}) is below $1\%$ for the parameter space of interest. The final expression we employ to find the percolation temperature is then
\begin{equation}
    e^{-G_*}\sqrt{\frac{2\pi}{G''_*}}=0.34\,.
    \label{eq:approx_percolation_final}
\end{equation}

When studying GW generation in supercooled FOPTs, it is fundamental to also check that the physical volume of the Universe in the false vacuum is shrinking at the percolation temperature~\cite{Athron:2023xlk,Athron:2023mer}. For this to be the case, one finds that the following inequality needs to be satisfied at $T_p$:
\begin{equation}
	T\frac{dI(T)}{dT}\Bigg|_{T_p}<-3\,.
    \label{eq:general_completion}
\end{equation}
Using the expression for $I(T)$ from Eq.~(\ref{eq:approx_percolation_final}) and noting that, from the definition of $G(T,T')$ in Eq.~(\ref{eq:function_exp_percolation}), only $f(T,T')$ depends on $T$, we rewrite Eq.~(\ref{eq:general_completion}) as
\begin{equation}
	\begin{split}
		\left[f(T,T_*)^{-1}\frac{df(T,T_*)}{dT} - \frac{1}{2G_*^{''}}\frac{d}{dT}\left(\frac{f'(T,T_*)^2}{f(T,T_*)^2}-\frac{f''(T,T_*)}{f(T,T_*)}\right)\right]\Bigg|_{T=T_p}< -\frac{3}{T_p I(T_p)}\,.
	\end{split}
	\label{eq:condition_fraction_false}
\end{equation}

\begin{figure}
    \includegraphics[width = 0.49\linewidth]{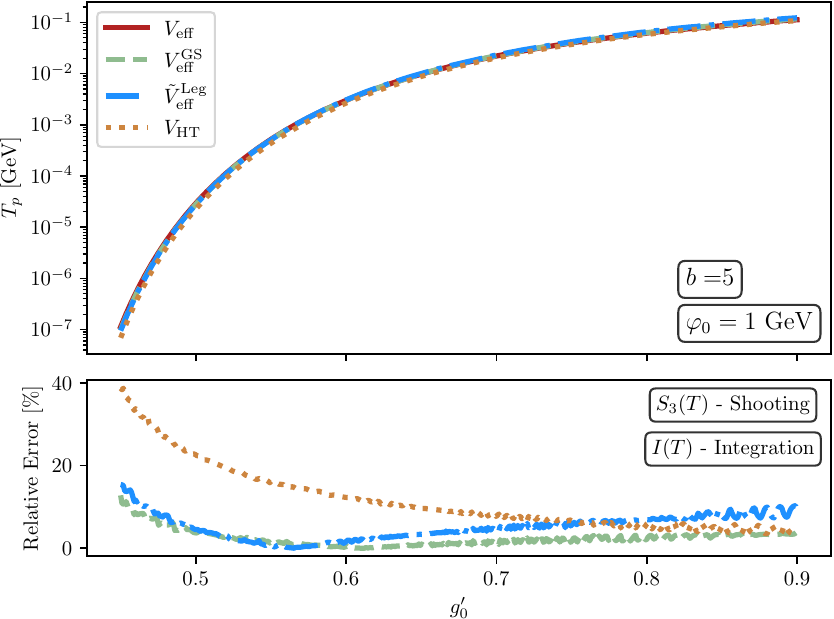}
    \includegraphics[width = 0.49\linewidth]{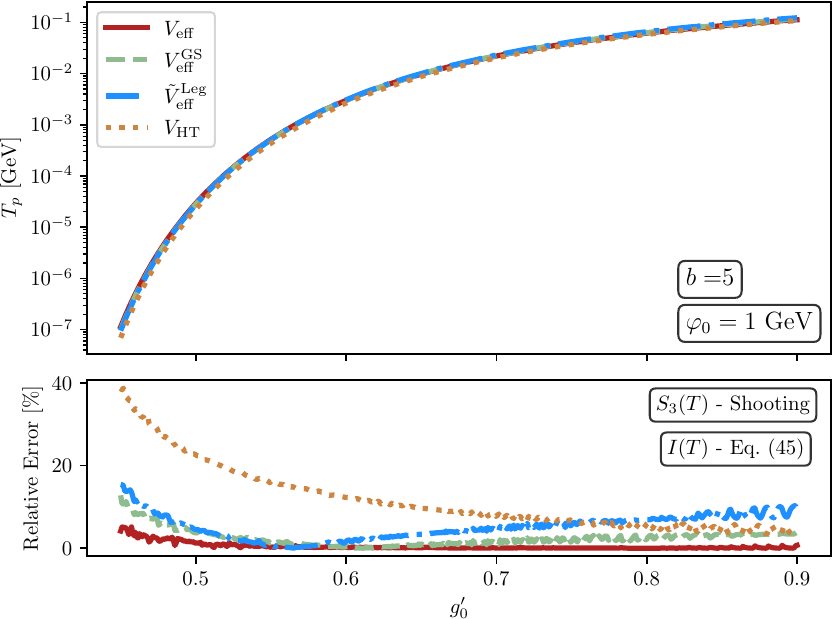}
    \includegraphics[width = 0.49\linewidth]{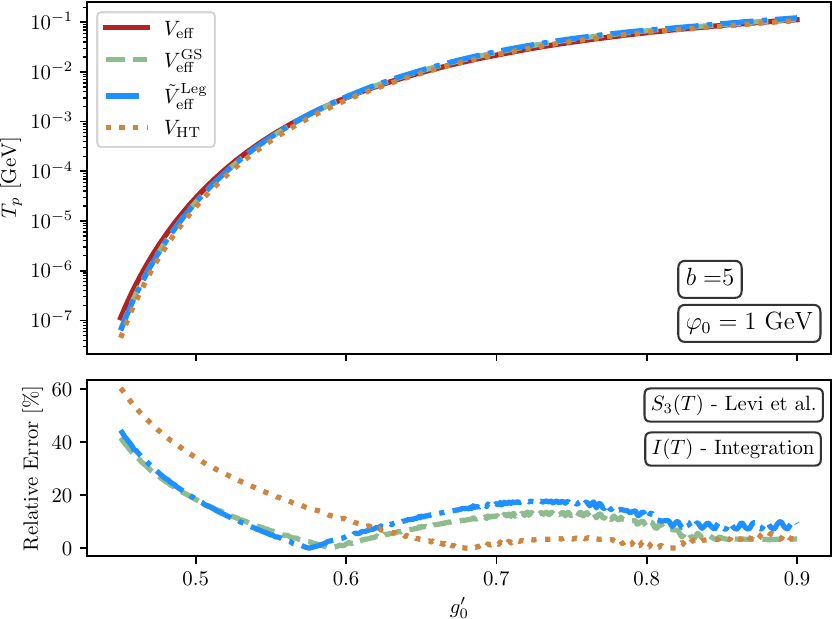}
    \includegraphics[width = 0.49\linewidth]{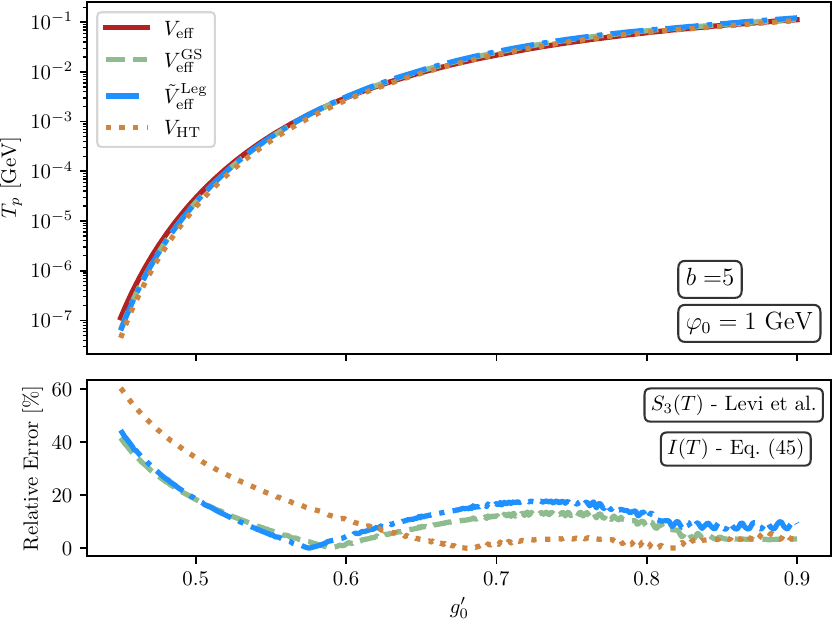}
    \caption{Comparison between different approximations for the computation of the percolation temperature. The red line corresponds to the full numerical computation in the plots showing $T_p$ vs $g'_0$. The different colored lines correspond to the same approximations for the effective potential as in Fig.~\ref{fig:comparison_actions}. In the upper panels we compute the action with a shooting method, while in the lower ones we use the action for the polynomial potentials from Ref.~\cite{Levi:2022bzt}. In the left panels the percolation temperature is found by integrating Eq.~(\ref{eq:percolation_full}) numerically. Finally, in the right panels we instead solve Eq.~(\ref{eq:approx_percolation_final}) to find $T_p$. The solid red line in the plot for the relative error compares the full solution found integrating against using Eq.~(\ref{eq:approx_percolation_final}). 
    }
    \label{fig:comparison_percolation_integration}
\end{figure}
We show the results for $T_p$ in Fig.~\ref{fig:comparison_percolation_integration}, together with the relative error between the full numerical solution and different approximations. In the left panels, we integrate $I(T)$ numerically, while in the right ones we use the analytical result from Eq.~(\ref{eq:approx_percolation_final}). In the upper panels, we compute $S_3(T)$ numerically~\cite{Wainwright:2011kj}, while in the lower ones we use the semi-analytical results from Eq.~(\ref{eq:fit_action})~\cite{Levi:2022bzt}. We use the same color code as before for all the plots showing the dependence of $T_p$ on $g'_0$, while fixing $\varphi_0=1\,\mathrm{GeV}$ and $b=5$. 

In order to highlight the quality of Eq.~(\ref{eq:approx_percolation_final}) for the determination of $T_p$, we point the reader to the solid red line shown in the upper right panel of Fig.~\ref{fig:comparison_percolation_integration} for the relative error as a function of $g'_0$. This line represents the relative error between the percolation temperature found using the full effective potential and a shooting method for $S_3$, either integrating numerically $I(T)$, or using instead Eq.~(\ref{eq:approx_percolation_final}). From this, we conclude that using Eq.~(\ref{eq:approx_percolation_final}) to find $T_p$ is always an excellent approximation, at most $\mathcal{O}(5\%)$ away from the full numerical result.

From Fig.~\ref{fig:comparison_percolation_integration} it is also clear that smaller values of the gauge coupling translate into larger supercooling, with percolation temperatures as low as $T_p\sim 10^{-7}$~GeV. To understand this fact analytically, we can study the dependence of $S_3$ from Eq.~(\ref{eq:fit_action}) on the Lagrangian parameters. Indeed, we expect $S_3/T\propto 1/g^{\prime 3}$, such that smaller gauge couplings tend to reduce the tunneling rate and thus delay the time at which 29\% of the Universe is already in the true vacuum. Obviously, such low percolation temperatures might face constraints from BBN and CMB~\cite{Bai:2021ibt,Xu:2025zsv}, but this goes beyond the scope of our work. 

Regarding the precision of the approximations introduced in Section~\ref{sec:Veff_poly}, it is clear in all panels that the inclusion of Daisy terms (shown by dashed green and dash-dotted blue lines) is fundamental for large supercooling, as neglecting these terms (dotted yellow line) introduces an error in the determination of $T_p$ more than twice as large. This can be seen in the upper panels where we compute $S_3$ numerically. For mild supercooling, i.e. larger values of $g'_0$, we nevertheless note that neglecting Daisy terms is not as critical, as the results for $T_p$ are comparable to those where we include Daisy contributions as polynomial terms. The reason is that the position and width of the potential barrier are inversely proportional to $g'_0$ (see Eq.~(\ref{eq:width_barrier}) for the width), such that larger values of the gauge coupling bring the barrier closer to the origin and consequently also the escape point $\varphi_{\mathrm{sc}}$ (cf. Fig.~\ref{fig:Comparison_HT}). In this region, Daisy contributions are well-described by a $\varphi$-independent term, which is irrelevant for the tunneling process. 

Next, we focus on the lower panels to discuss the impact of Eq.~(\ref{eq:fit_action}) to compute $S_3$. As previously discussed in Fig.~\ref{fig:comparison_actions}, the semi-analytical action tends to underestimate the value of $S_3$, such that there is a cancellation between the error introduced by the use of Eq.~(\ref{eq:fit_action}) and that from neglecting Daisy terms. This explains the better results for $V_{\mathrm{HT}}$ in the lower panels of Fig.~\ref{fig:comparison_percolation_integration} with respect to the cases where we explicitly include Daisy resummation. Nevertheless, this is not the consequence of a controlled set of approximations, but just the result of two opposite effects canceling each other. Instead, we note that for large supercooling this effect is no longer there, and we consistently find again that including Daisy contributions (shown as dash-dotted blue and dashed green lines) is necessary to reduce the overall relative error in $T_p$. 

Finally, comparing the upper and lower panels, in which the only difference is whether we compute $S_3$ numerically or with the semi-analytical result from Ref.~\cite{Levi:2022bzt}, we conclude that the larger source of error in the different approximations arises from the use of Eq.~(\ref{eq:fit_action}), more than doubling the error committed in computing $T_p$ when including Daisy terms. Given that we find that Eq.~(\ref{eq:condition_fraction_false}) is not satisfied for $g'_0\lesssim 0.5$, such that there are no successful FOPTs below $g'_0=0.5$, we conclude that the approximations we proposed in Section~\ref{sec:Veff_poly}, and the one used to compute $T_p$ through Eq.~(\ref{eq:approx_percolation_final}), introduce at most a $\mathcal{O}(5\%)$ error (see the two upper-right panels). Further adopting the semi-analytical action from Ref.~\cite{Levi:2022bzt} increases this up to $\mathcal{O}(20\%)$, as shown in the lower-right panels. 

\subsection{Completion temperature}
It is also important to assess whether the phase transition completes, i.e. that there are no remaining fraction of the Universe in the false vacuum. This can be quantified by requiring that the fraction of the Universe in the false vacuum is negligible,
\begin{equation}
    P_f(T_\mathrm{comp})=\epsilon\,,
\end{equation}
with $\epsilon\ll 1$~\cite{Athron:2023xlk}. For simplicity, and since its distinction is not relevant for our analysis, we we identify the completion temperature with the percolation one, such that $\epsilon = 0.29$. Nevertheless, thanks to our semi-analytical estimate for the false vacuum fraction, Eq.~(\ref{eq:approx_percolation_final}) can be straightforwardly generalized to determine $T_{\mathrm{comp}}$ for any choice of $\epsilon$.

\section{Gravitational wave spectrum}\label{sec:results}
Once the bounce action has been determined as a function of the temperature, we can obtain the percolation temperature as previously discussed, but also the FOPT inverse duration, which is another input parameter necessary for the prediction of the GW spectrum. We refer the reader to Refs.~\cite{Athron:2023xlk,Caprini:2024hue,Ellis:2019oqb} for details on the GW templates and subtleties arising when trying to go from the particle physics model to the thermal parameters controlling the generation of GWs, and just summarize here the main quantities that need to be taken into account in the analysis of GW generation in FOPTs.
\begin{figure}
    \begin{center}
        \includegraphics[width = 0.49\linewidth]{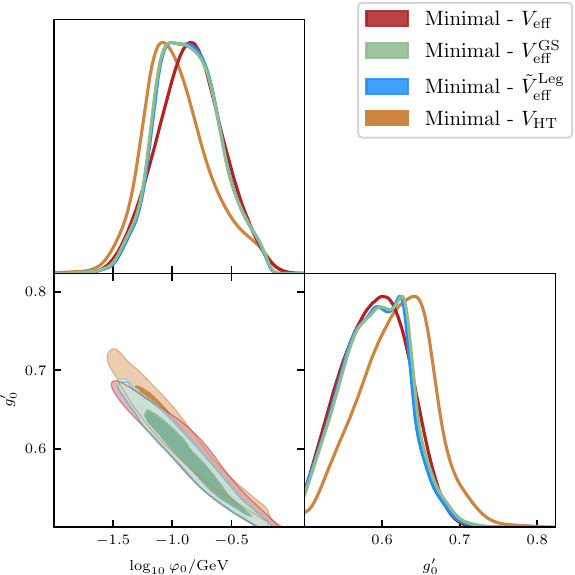}
        \includegraphics[width = 0.49\linewidth]{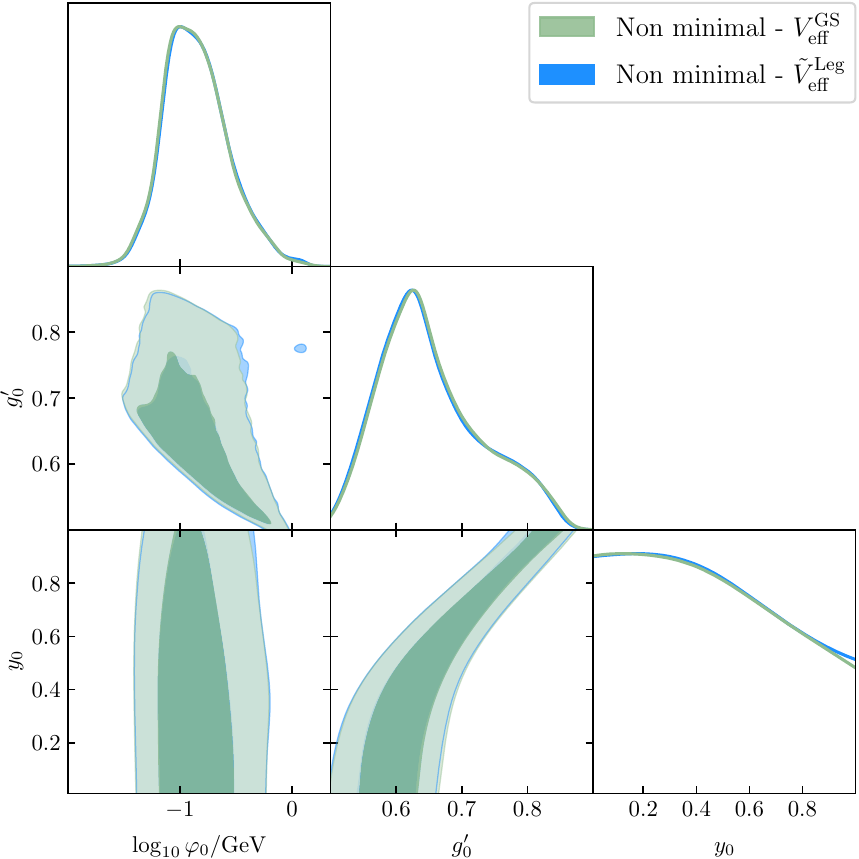}
        \caption{Posterior distributions for the allowed parameter space compatible with the NANOGrav signal, found using \texttt{PTArcade}~\cite{Mitridate:2023oar}. We show the 68\% and 95\% posterior distribution with dark and lighter hues, respectively. The left panel corresponds to the minimal scenario we study in Section~\ref{sec:Veff_poly}. The red region corresponds to the full numerical computation. Green and blue regions correspond to our full pipeline of approximations, using the Gram-Schmidt and Legendre approximations for Daisy terms, respectively, while the yellow one is the result found neglecting Daisy terms. The right panel shows the posterior distribution in a non-minimal model, specified in Appendix~\ref{app:running_non-minimal}, using the polynomial potentials $V_{\mathrm{eff}}^{\mathrm{GS}}$ and $\tilde{V}_{\mathrm{eff}}^{\mathrm{Leg}}$.}
        \label{fig:PTArcade_minimal}
    \end{center}
\end{figure}

\begin{itemize}
	\item Percolation temperature $T_p$:
    
	As already discussed extensively, this is the temperature at which 29\% of the Universe is already in the true vacuum and that at which numerical simulations find that GWs are produced. In our approach we compute it using Eq.~(\ref{eq:approx_percolation_final}).
    
	\item FOPT strength $\alpha$:

	While $\alpha$ generally depends on the effective potential at finite temperature, given that we focus on supercooled FOPT, we can neglect temperature corrections and simply study the effective potential in vacuum, setting $\mu^2=\mu_0^2= m_{Z',0}^2$~\cite{Kierkla:2023von}. This allows us to compute the strength of the FOPT directly from the $T=0$ potential very precisely at the percolation temperature $T_p$
    \begin{equation}
        \begin{split}
            \alpha=&\frac{1}{\rho_{\mathrm{rad}}(T_p)}\left[ V_{\mathrm{eff}}(0,T)-V_{\mathrm{eff}}(\varphi_0,T)-\frac{T}{4}\frac{\partial(V_{\mathrm{eff}}(0,T)-V_{\mathrm{eff}}(\varphi_0,T))}{\partial T}\right]\Bigg|_{T_p} \\
            \simeq & \frac{1}{\rho_{\mathrm{rad}}(T_p)}\left(-\lambda(t)+\frac{5 g'^4(t)}{32\pi^2}\right)\Bigg|_{\mu^2_0}\frac{\varphi^4_0}{4} = \frac{1}{\rho_{\mathrm{rad}}(T_p)}\frac{3g_0'^4}{128\pi^2}\varphi_0^4\,,
        \end{split}
        \label{eq:alpha_RSB}
    \end{equation}
    where the last equality follows from Eq.~(\ref{eq:relation_lambda_gprime}).
    
	\item Reheating temperature $T_{\mathrm{re}}$:
    
	This is related to both the percolation temperature and the released energy by the FOPT, and it is the temperature of the Universe after the FOPT is over and the scalar field has deposited all its energy in the plasma. It is approximately given by
	\begin{equation}
		T_{\mathrm{re}}\simeq (1+\alpha)^{1/4}T_p\,.
	\end{equation}
    
	\item Inverse duration of the FOPT $\beta/H$:

	Approximating the nucleation rate near $T_p$ as $\Gamma(t) \simeq \Gamma_p e^{\beta (t-t_p)}$, with $t_p$ being the corresponding time to the percolation temperature, the inverse duration is given by
	\begin{equation}
		\frac{\beta}{H(T_p)}\simeq T\frac{d}{dT}\left(\frac{S_3(T)}{T}\right)\Bigg|_{T_p}\left[1-\frac{3}{2}\left(\frac{S_3(T_p)}{T_p}\right)^{-1}\right]-4\,.
		\label{eq:beta}
	\end{equation}
    We adopt Eq.~(\ref{eq:fit_action}) to compute the derivative of $S_3(T)$ in $\beta/H$, and highlight that the second term in the square brackets is subdominant. We note the subtleties regarding the description of the nucleation rate only in terms of an exponential~\cite{Athron:2023xlk}, but these are beyond the scope of our work.
    
    \item Bubble wall velocity $v_w$:
    
    We assume that the bubble wall quickly reaches relativistic velocities in the runaway regime, which is expected in supercooled FOPTs. Therefore, we set $v_w\rightarrow 1$. However, the NLO pressure terms from gauge bosons can prevent the bubble walls from accelerating indefinitely~\cite{Bodeker:2009qy,Bodeker:2017cim,Gouttenoire:2021kjv}, such that they reach a steady state velocity given by the Lorentz factor
    \begin{equation}
        \gamma_{\mathrm{eq}}=\frac{\Delta V_{\mathrm{eff}} - \mathcal{P}_{\mathrm{LO}}}{\mathcal{P}_{\mathrm{NLO}}}\,.
    \end{equation}
    In the runaway regime we have $\Delta V_{\mathrm{eff}}>\mathcal{P}_{\mathrm{LO}}$, and $\mathcal{P}_{\mathrm{(N)LO}}$ are the (next-to-) leading order pressure terms whose expressions can be found, for example, in Ref.~\cite{Gouttenoire:2021kjv}. Additionally, we need to compare $\gamma_{\mathrm{eq}}$ to the Lorentz factor the walls would reach in the absence of $\mathcal{P}_{\mathrm{NLO}}$, $\gamma_*$, in order to correctly understand the fraction of energy from the FOPT that goes into the production of GWs from bubble collisions, sound waves, and turbulence, respectively. Following Ref.~\cite{Gouttenoire:2021kjv}, we compute 
    \begin{equation}
        \gamma_* = \frac{R_p}{3 R_0}\,,
    \end{equation}
    with $R_p$ the bubble radius at collision which we compute following the approximation advocated in Ref.~\cite{Caprini:2024hue}
    \begin{equation}
        R_p\equiv \frac{(8\pi)^{1/3} \mathrm{max}(v_w,c_s)}{\beta}\,,
    \end{equation}
    with $c_s=1/\sqrt{3}$ being the sound speed in the plasma, and where
    \begin{equation}
        R_0=\left(\frac{3S_3(T_p)}{2\pi \Delta V_{\mathrm{eff}}}\right)^{1/3}
    \end{equation}
    represents the initial bubble radius at nucleation. With these ingredients we follow Ref.~\cite{Ellis:2019oqb} in order to compute the efficiency factors for each GW source. 
\end{itemize}

After determining these thermal parameters, the GW spectrum can be computed using the templates of Ref.~\cite{Caprini:2024hue} for each GW source, including the efficiency factors as prescribed in Ref.~\cite{Ellis:2019oqb}. In order to assess the validity of our approximations to find the viable regions of parameter space compatible with the NANOGrav signal~\cite{NANOGrav:2023gor,NANOGrav:2023hvm}, we use \texttt{PTArcade}~\cite{Mitridate:2023oar} to obtain the $1\,(2)\,\sigma$ Bayesian posterior distributions for $\varphi_0$ and $g'_0$, adopting the default configuration \texttt{ceffyl}~\cite{lamb2023rapid} and neglecting contributions from supermassive black hole binaries. For convenience, we summarize in Table~\ref{tab:summary} the minimal set of steps required to adopt our results for a generic particle-physics model.

\begingroup
\begin{table}[h]
\centering
\begin{tabular}{p{2.8cm}|p{3.9cm}|p{4.6cm}|p{3.0cm}}
\textbf{Step} 
  & \textbf{Approximation} 
  & \textbf{Assumptions} 
  & \textbf{Relative error} \\
\hline
$V_{\rm Daisy}$
  & Polynomial projection 
    ($V_{\rm Daisy}^{\rm GS(Leg)}$)
  & $\varphi/T \lesssim b \sim 5$
  & $\mathcal{O}(1\%)$ in $V_{\rm eff}$ \\
\hline
$J_{b(f)}(T)$
  & HT expansion
  & $m^2_{b(f)}\propto\varphi^2$,\;
    $m_{b(f)}/T \ll 1$
  & $\mathcal{O}(2\%)$ in $S_3$ \\
\hline
Polynomial $V_{\rm eff}$
  & $V_{\rm HT} + V_{\rm Daisy}^{\rm GS(Leg)}$,
    Eqs.~(\ref{eq:Veff_GS_final})--(\ref{eq:Veff_Leg_final})
  & Both conditions above
  & $\mathcal{O}(2\%)$ in $S_3$ \\
\hline
Renormalization scale
  & $\mu = \max[m_{Z'}(\varphi),\, \pi T]$
  & Best agreement with 
    3D EFT~\cite{Christiansen:2025xhv}
  & See Ref.~\cite{Christiansen:2025xhv} \\
\hline
Bounce action
  & Fit from Ref.~\cite{Levi:2022bzt},
    Eq.~(\ref{eq:fit_action})
  & Polynomial $V_{\rm eff}$,\;
    $g'_0 \lesssim 1$
  & $\mathcal{O}(5\%)$ in $S_3$
    \textbf{(dominant source)} \\
\hline
Percolation temperature
  & Laplace approximation, 
    Eq.~(\ref{eq:approx_percolation_final})
  & Single maximum in integrand 
    of Eq.~(\ref{eq:percolation_full})
  & ${\lesssim}\,\mathcal{O}(1\%)$ in $T_p$
    (independent of rows above) \\
\end{tabular}
\caption{Summary of approximations and their related uncertainties for the full pipeline. Errors are relative to the full numerical computation for the U(1)$'$ scale-invariant model with $\varphi_0 = 1$~GeV. The percolation-temperature error in the last row is independent of the $V_{\rm eff}$ and $S_3$ approximations above it.
}
\label{tab:summary}
\end{table}
\endgroup

We show the results in the left panel of Fig.~\ref{fig:PTArcade_minimal}, comparing the full computation, shown as a red region, with the full pipeline of approximations using the Gram-Schmidt (green) and Legendre (blue) decompositions for Daisy contributions. We additionally show the result of neglecting completely Daisy contributions as a yellow region. While the precise determination of the GW spectrum is indeed sensitive to a variation of the thermal parameters, we conclude that our approximations are precise enough for phenomenological purposes in models which are not strongly constrained. The results from the full computations or for the Gram-Schmidt and Legendre approaches shown in Fig.~\ref{fig:PTArcade_minimal} could now be used to study the reach of laboratory experiments in probing this parameter space, with little impact from the small differences in the preferred parameter space from PTAs. While this is also true even when using only the HT approach and neglecting Daisy contributions~\cite{Lewicki:2024xan}, we went beyond these by including Daisy terms as polynomials and simplifying the approach adopted to find the percolation temperature, improving the agreement with a full numerical analysis. 

An important advantage of our proposed approximations is that they render the scan of complete parameter spaces feasible: we go from running the necessary Markov Chain Monte Carlo (MCMC) algorithm in \texttt{PTArcade} for weeks to doing so in hours when exploiting these approximations. This allows to assess the viability of a given scenario to produce a signal in GW experiments, and its complementarity with laboratory searches without resorting to heavy numerical computations. Additionally, writing the effective potential as a polynomial allowed for various insights on the dependence of the results of the couplings and masses of the NP scenario under study. 

One can now apply this methodology to study more involved scenarios predicting supercooled FOPTs, assessing their viability and allowed parameter space. As an example, we consider the same $U(1)^{\prime}$ scale-invariant scenario, but including fermions. The most relevant consequence from the introduction of fermions is that now the Yukawa terms as specified in Eq.~(\ref{eq:lag_non-minimal}) in Appendix~\ref{app:running_non-minimal} also play a role in the radiative symmetry breaking, such that the following condition
\begin{equation}
    \lambda_0=\frac{1}{16\pi^2}\left[g_0^{\prime4}-y_0^4\left(1-\log{\frac{y_o^2}{2g_0^{\prime2}}}\right)\right]\,,
    \label{eq:RSB_non-minimal_0}
\end{equation}
needs to be satisfied at the renormalization scale $\mu_0=m_{Z',0}$. Given that fermion fields do not introduce new Daisy terms in the effective potential, we can directly use all our previous results by only modifying $V_{\mathrm{HT}}$, as well as taking into account the changes in the RG evolution with the new couplings involved. These modifications are summarized in Appendix~\ref{app:running_non-minimal}. The results are shown in the right panel of Fig.~\ref{fig:PTArcade_minimal}, using now solely our full pipeline of approximations with $V_{\mathrm{eff}}^{\mathrm{GS}}$ (green region) and $\tilde{V}_{\mathrm{eff}}^{\mathrm{Leg}}$ (blue one). As can be realized, the inclusion of the fermion Yukawa coupling opens up the allowed values of $g'_0$ explaining the NANOGrav signal at $1\sigma$, while finding a strong correlation between $y_0$ and $g'_0$. The latter is a consequence of Eq.~(\ref{eq:RSB_non-minimal_0}), and implies a hierarchy of masses between the dark particles, that could have strong phenomenological consequences to be searched for. 

\section{Conclusions}\label{sec:conclusions}
In this work we have studied the possibility to perform the full pipeline of computations pertaining to the prediction of GW production during cosmological FOPTs semi-analytically in classically scale-invariant scenarios. While the starting point is common to previous studies, namely the HT expansion of the effective potential allowing to write it down in terms of a polynomial, we have gone beyond this by consistently introducing Daisy contributions as well. To the best of our knowledge, the latter possibility had never been investigated in detail, even though it is known that Daisy contributions play a critical role in the HT limit of the effective potential.

Once the effective potential is described as a polynomial, the semi-analytical results from Ref.~\cite{Levi:2022bzt} for $S_3$ can be exploited, avoiding to solve numerically the equations of motion for the bounce in every point of the parameter space. A crucial point we also investigated is the inclusion of the RG evolution, given that the natural choice for the renormalization scale at small field values, which is the relevant region for tunneling, is proportional to the temperature of the Universe. While the proportionality constant is somewhat arbitrary from the perspective of the $4D$ theory, we set it to the value for which the $4D$ effective potential agrees the most with the corresponding $3D$-NLO result~\cite{Christiansen:2025xhv}. The latter represents the state-of-the-art approach for the study of the effective potential in FOPTs. 

Next, and independently from previous approximations on the effective potential or $S_3$, we have obtained an analytical estimate for the fraction of the Universe in the false vacuum, greatly simplifying the computation of the percolation temperature. This is important given that, while the nucleation temperature is usually adopted as a proxy for the moment at which GWs are produced, it is instead the percolation temperature the one that sets the time for GW production, as found in numerical simulations. Instead of relying on computationally expensive numerical integrations, we performed the full double integral analytically, allowing to reduce the computation of the percolation temperature from a fully numerical double integration to a root-finding routine, similar to what is done for the nucleation temperature. 

Our method allowed us also to assess the error introduced in each approximation with respect to the full numerical computation. For the relevant region of parameter space, we found that our description of the effective potential including Daisy terms introduces a $\mathcal{O}(2\%)$ error in the determination of the bounce action, being even smaller for large supercooling. This translates into at most a $\mathcal{O}(5\%)$ error in the percolation temperature when compared to a full numerical determination, in contrast to the $\mathcal{O}(20\%)$ error in the same region of parameter space when neglecting Daisy terms altogether. Nevertheless, we note that not including Daisy resummation does not represent such a bad approximation to the full effective potential for small supercooling. In general, we find that the semi-analytical result on $S_3$ for polynomial potentials~\cite{Levi:2022bzt} represents the biggest source of error with respect to the full numerical computation in our approach. Instead, the analytical estimation of the fraction of the Universe in the false vacuum we proposed, simplifying the determination of the percolation temperature, introduces at most a $\mathcal{O}(1\%)$ error, resulting in an excellent estimation of the percolation temperature regardless of the method adopted to compute $S_3$.

We applied our results to the most minimal scenario known to account for the SGWB observed in PTAs, namely a classically scale-invariant model with a new $U(1)^\prime$ gauge symmetry, the associated gauge boson and a scalar that breaks it spontaneously. We find that indeed our approximations can be employed not only to avoid the use of computationally intensive methods, but also to gain some insights on the impact of different particle masses and couplings in the final results. We obtain very good agreement between the allowed parameter space found using the full numerical procedure or our pipeline of approximations, while doing so in much less time in the latter case. Finally, profiting from the relatively cheap computational cost of the use of our approximations, we have also studied a second scenario including fermions. Their inclusion impacts the radiative symmetry breaking, and enlarges the allowed parameter space while potentially offering a very rich phenomenology that could be searched for in experiments. 

All in all, these results could open the window to efficient parameter scans compatible with GW signals that could then be used for phenomenological purposes, enhancing the synergies between laboratory experiments and potential GW observations.

\section*{Acknowledgments}
The authors thank Filippo Sala for very fruitful discussions and encouragement, as well as Alessia Musumeci and Jacopo Nava for useful comments on the manuscript and discussions on the gravitational wave templates. They also thank Giacomo Ferrante and Joonas Hirvonen for very enlightening commenvonts on the tunneling action. SRA warmly thanks Javier Orts for useful discussions about the percolation temperature. This work has been partly funded by the European Union under the Horizon Europe’s Project: 101201278 – DarkSHunt - ERC - 2024 ADG. Views and opinions expressed are however those of the author(s) only and do not necessarily reflect those of the European Union or the European Research Council Executive Agency. Neither the European Union nor the granting authority can be held responsible for them.

\appendix
\section{Polynomial coefficients for the Daisy contribution}\label{app:Daisy_poly_coefficients}
We give the explicit expressions for the polynomial coefficients obtained when decomposing the Daisy contribution from Eq.~(\ref{eq:term_Daisy_expansion}) in terms of orthogonal polynomials. For the decomposition in terms of Legendre polynomials in an interval $x\in[0,b]$, and remembering that we parametrize the thermal and vacuum masses of the relevant boson as $\Pi_b(T)=C_b g'^2 T^2$ (assuming $C_b>0$) and $m_b^2(\varphi)=g'^2\varphi^2$ respectively, we find
    \begin{align}
            c^{\mathrm{Leg}}_0 = (8 b C_b)^{-1} \Bigg[&3 C_b^{3/2} \Big[\log \left(-\sqrt{b^2+4 C_b}+b+2\sqrt{C_b}\right) \nonumber\\
            +&\log \left(2 \sqrt{b^2+C_b}-\sqrt{b^2+4C_b}+b\right) \nonumber\\
            -&\log \left(\left(\sqrt{b^2+4 C_b}+b-2\sqrt{C_b}\right) \left(-2 \sqrt{b^2+C_b}+\sqrt{b^2+4C_b}+b\right)\right)\Big] \nonumber\\
            +&5 b \sqrt{C_b\left(b^2+C_b\right)}+2 b^3\sqrt{\frac{b^2+C_b}{C_b}}\Bigg]\,,
            \label{eq:c0_Legendre}\\
            c^{\mathrm{Leg}}_1 = 3(40 b^2C_b)^{-1} \Bigg[&-15 b C_b^{3/2} \log \left(-\sqrt{b^2+4 C_b}+b+2\sqrt{C_b}\right) \nonumber\\
            -&15 b C_b^{3/2} \log \left(2\sqrt{b^2+C_b}-\sqrt{b^2+4 C_b}+b\right) \nonumber\\
            +&15 b C_b^{3/2}\log \left(\sqrt{b^2+4 C_b}+b-2 \sqrt{C_b}\right) \nonumber\\
            +&15 bC_b^{3/2} \log \left(-2 \sqrt{b^2+C_b}+\sqrt{b^2+4C_b}+b\right) \nonumber\\
            +&16 \sqrt{C_b^3 \left(b^2+C_b\right)}+7b^2 \sqrt{C_b \left(b^2+C_b\right)}+6 b^4\sqrt{\frac{b^2+C_b}{C_b}}-16 C_b^2\Bigg]\,,
            \label{eq:c1_Legendre}\\
            c^{\mathrm{Leg}}_2 = (16 b^3C_b)^{-1}\Bigg[&120 b^2 C_b^{3/2} \log \left(2 \sqrt{b^2+C_b}-\sqrt{b^2+4C_b}+b\right) \nonumber\\
            -&120 b^2 C_b^{3/2} \log \left(\sqrt{b^2+4C_b}+b-2 \sqrt{C_b}\right) \nonumber\\
            -&120 b^2 C_b^{3/2} \log\left(-2 \sqrt{b^2+C_b}+\sqrt{b^2+4 C_b}+b\right) \nonumber\\
            -&60C_b^{5/2} \log \left(2 \sqrt{b^2+C_b}-\sqrt{b^2+4C_b}+b\right) \nonumber\\
            +&60 C_b^{5/2} \log \left(\sqrt{b^2+4C_b}+b-2 \sqrt{C_b}\right) \nonumber\\
            +&60 C_b^{5/2} \log \left(-2\sqrt{b^2+C_b}+\sqrt{b^2+4 C_b}+b\right) \nonumber\\
            -&66 b\sqrt{C_b^3 \left(b^2+C_b\right)}+4 b^5\sqrt{\frac{b^2+C_b}{C_b}} \nonumber\\
            -&15 \sqrt{C_b} \left(b^4-8b^2 C_b+4 C_b^2\right) \log \left(-\sqrt{b^2+4 C_b}+b+2\sqrt{C_b}\right) \nonumber\\
            +&15 \sqrt{C_b} \left(b^4-6 b^2 C_b+2C_b^2\right) \sinh ^{-1}\left(\frac{b}{\sqrt{C_b}}\right) \nonumber\\
            -&15b^4 \sqrt{C_b} \log \left(2 \sqrt{b^2+C_b}-\sqrt{b^2+4C_b}+b\right) \nonumber\\
            +&15 b^4 \sqrt{C_b} \log \left(\sqrt{b^2+4C_b}+b-2 \sqrt{C_b}\right) \nonumber\\
            +&15 b^4 \sqrt{C_b} \log\left(-2 \sqrt{b^2+C_b}+\sqrt{b^2+4 C_b}+b\right) \nonumber\\
            -&2 b^3\sqrt{C_b \left(b^2+C_b\right)}+96 b C_b^2\Bigg]\,,
            \label{eq:c2_Legendre}\\
            c^{\mathrm{Leg}}_3 = (80 b^4 C_b)^{-1}\Bigg[&2100 b C_b^{5/2} \log \left(2 \sqrt{b^2+C_b}-\sqrt{b^2+4C_b}+b\right) \nonumber\\
            -&2100 b C_b^{5/2} \log \left(\sqrt{b^2+4C_b}-2 \sqrt{C_b}\right) \nonumber\\
            -&2100 b C_b^{5/2} \log\left(\frac{\sqrt{b^2+4 C_b}+b-2 \sqrt{C_b}}{\sqrt{b^2+4C_b}-2 \sqrt{C_b}}\right) \nonumber\\
            -&2100 b C_b^{5/2} \log\left(-2 \sqrt{b^2+C_b}+\sqrt{b^2+4 C_b}+b\right) \nonumber\\
            -&640\sqrt{C_b^5 \left(b^2+C_b\right)}+614 b^2 \sqrt{C_b^3\left(b^2+C_b\right)} -1344 b^2 C_b^2 \nonumber\\
            +&4 b^6\sqrt{\frac{b^2+C_b}{C_b}} 
            -175 b^5 \sqrt{C_b} \log\left(2 \sqrt{b^2+C_b}-\sqrt{b^2+4 C_b}+b\right) \nonumber\\
            +&175 b^5\sqrt{C_b} \log \left(\sqrt{b^2+4 C_b}-2\sqrt{C_b}\right) \nonumber\\
            +&175 b^5 \sqrt{C_b} \log\left(\frac{\sqrt{b^2+4 C_b}+b-2 \sqrt{C_b}}{\sqrt{b^2+4C_b}-2 \sqrt{C_b}}\right) \nonumber\\
            +&175 b^5 \sqrt{C_b} \log\left(-2 \sqrt{b^2+C_b}+\sqrt{b^2+4 C_b}+b\right) \nonumber\\
            -&35 b\sqrt{C_b} \left(5 b^4-4 b^2 C_b-60 C_b^2\right) \log\left(-\sqrt{b^2+4 C_b}+b+2 \sqrt{C_b}\right) \nonumber\\
            +&175 b\sqrt{C_b} \left(b^4-2 b^2 C_b-6 C_b^2\right) \sinh^{-1}\left(\frac{b}{\sqrt{C_b}}\right)-2 b^4 \sqrt{C_b\left(b^2+C_b\right)} \nonumber\\
            +&140 b^3 C_b^{3/2} \log \left(2\sqrt{b^2+C_b}-\sqrt{b^2+4 C_b}+b\right) \nonumber\\
            +&175 b^3C_b^{3/2} \log \left(\sqrt{b^2+4 C_b}-2\sqrt{C_b}\right) \nonumber\\
            -&315 b^3 C_b^{3/2} \log \left(\sqrt{b^2+4C_b}+b-2 \sqrt{C_b}\right) \nonumber\\
            +&175 b^3 C_b^{3/2} \log\left(\frac{\sqrt{b^2+4 C_b}+b-2 \sqrt{C_b}}{\sqrt{b^2+4C_b}-2 \sqrt{C_b}}\right) \nonumber\\
            -&140 b^3 C_b^{3/2} \log\left(-2 \sqrt{b^2+C_b}+\sqrt{b^2+4 C_b}+b\right)+640C_b^3\Bigg]\,,
            \label{eq:c3_Legendre}\\
            c^{\mathrm{Leg}}_4 = \frac{9}{64 b^5\sqrt{C_b}}\Bigg[&b \Big[256 b^2 C_b^{3/2}+407 C_b^2\sqrt{b^2+C_b}-120 b C_b^2 \log \left(\sqrt{b^2+4C_b}+b-2 \sqrt{C_b}\right) \nonumber\\ 
            -&120 b C_b^2 \log \left(-2\sqrt{b^2+C_b}+\sqrt{b^2+4 C_b}+b\right)-82 b^2 C_b\sqrt{b^2+C_b}\nonumber\\
            -&30 b^5 \log \left(\left(\sqrt{b^2+4 C_b}+b-2\sqrt{C_b}\right)\left(-2\sqrt{b^2+C_b}+\sqrt{b^2+4 C_b}+b\right)\right) \nonumber\\
            +&261 b^3 C_b\log \left(\left(\sqrt{b^2+4 C_b}+b-2 \sqrt{C_b}\right)\left(-2 \sqrt{b^2+C_b}+\sqrt{b^2+4C_b}+b\right)\right) \nonumber\\
            +&3 \left(10 b^5-87 b^3 C_b+40 bC_b^2\right) \log \left(\left(-\sqrt{b^2+4 C_b}+b+2\sqrt{C_b}\right)\right) \nonumber\\ 
            +&3 \left(10 b^5-87 b^3 C_b+40 bC_b^2\right) \log \left(2 \sqrt{b^2+C_b}-\sqrt{b^2+4C_b}+b\right)-512 C_b^{5/2}\Big] \nonumber\\
            -&15 \left(2 b^6-19 b^4C_b+32 b^2 C_b^2-7 C_b^3\right) \sinh^{-1}\left(\frac{b}{\sqrt{C_b}}\right)\Bigg]\,.
            \label{eq:c4_Legendre}
\end{align}
The Daisy contribution to the effective potential is then given by
\begin{equation}
	V_{\mathrm{Daisy}}^{\mathrm{Leg}}(\varphi,T)=V_{\mathrm{Leg},\,0} + \nu^3_{\mathrm{Leg}} \varphi + \frac{1}{2}m^2_{\mathrm{Leg}}\varphi^2 - \frac{1}{3}K_{\mathrm{Leg}}\varphi^3-\frac{1}{4}\lambda_{\mathrm{Leg}}\varphi^4\,,
\end{equation}
with the following polynomial coefficients:
\begin{equation}
	\begin{split}
			V_{\mathrm{Leg},0} =& -\frac{C_b^{3/2}g'^3T^4 (c^{\mathrm{Leg}}_0-c^{\mathrm{Leg}}_1+c^{\mathrm{Leg}}_2-c^{\mathrm{Leg}}_3+c^{\mathrm{Leg}}_4)}{12 \pi }\,,\\
		\nu^3_{\mathrm{Leg}} =& -\frac{C_b^{3/2}g'^3T^3 (c^{\mathrm{Leg}}_1-3 c^{\mathrm{Leg}}_2+6 c^{\mathrm{Leg}}_3-10 c^{\mathrm{Leg}}_4)}{6\pi  b}\,,\\
		m^2_{\mathrm{Leg}} = & -\frac{C_b^{3/2}g'^3T^2 (c^{\mathrm{Leg}}_2-5 c^{\mathrm{Leg}}_3+15 c^{\mathrm{Leg}}_4)}{\pi  b^2}\, ,\\
		K_{\mathrm{Leg}} =& \frac{g'^3 T \left(20 C_b^{3/2} (c^{\mathrm{Leg}}_3-7 c^{\mathrm{Leg}}_4)- b^3\right)}{4 \pi b^3}\,,\quad
		\lambda_{\mathrm{Leg}} =  \frac{70 C_b^{3/2}g'^3c^{\mathrm{Leg}}_4}{3 \pi  b^4} \,.
	\end{split}
	\label{eq:coefficients_Daisy_Leg}
\end{equation}
Given that this procedure introduces a tadpole that needs to be removed in order to use previous results on $S_3$, we shift the field as $\varphi\rightarrow \tilde{\varphi}-w$ and find the shifts that solve Eq.~(\ref{eq:implicit_cubic_tadpole}). The solution is given by\footnote{Note that there are three possible solutions, but we take the one with smallest absolute value}
\begin{equation}
    w_k=-\frac{1}{3(\lambda+\lambda_{\mathrm{Leg}})}\left[-(K+K_{\mathrm{Leg}})+\xi^kC+\frac{\Delta_0}{\xi^kC}\right]\,
\end{equation}
with $k=0,\,1,\,2$, and where we have defined
\begin{equation}
    \begin{split}
        \Delta_0 = &(K+K_{\mathrm{Leg}})^2+3(\lambda+\lambda_{\mathrm{Leg}})(m^2+m^2_{\mathrm{Leg}})\,,\\
        \Delta_1 = &- 2(K+K_{\mathrm{Leg}})^3-9(\lambda+\lambda_{\mathrm{Leg}})(K+K_{\mathrm{Leg}})(m^2+m^2_{\mathrm{Leg}})\\
        &+27(\lambda+\lambda_{\mathrm{Leg}})^2\nu^3_{\mathrm{Leg}}\,,\\
        C=&\left(\frac{\Delta_1\pm\sqrt{\Delta_1^2-4\Delta_0^3}}{2}\right)^{1/3}\,,\\
        \xi=&\frac{-1+\sqrt{-3}}{2}\,.
    \end{split}
    \label{eq:definitions_cubic_sol}
\end{equation}
The sign of the square root in the third line in Eq.~(\ref{eq:definitions_cubic_sol}) is arbitrary. In the case in which one choice of sign makes $C=0$, the other option should be taken.

Instead, for the Gram-Schmidt decomposition we find
\begin{align}
		c_2^{\mathrm{GS}}=\frac{21 C_b^{5/2}}{32 b^7}\Bigg[&12 \left(\frac{b}{\sqrt{C_b}}\right)^5\sqrt{\left(\frac{b}{\sqrt{C_b}}\right)^2+1}-32\left(\frac{b}{\sqrt{C_b}}\right)^5 -512\left(\frac{b}{\sqrt{C_b}}\right)\nonumber\\
		-&210\left(\frac{b}{\sqrt{C_b}}\right)^2\sinh^{-1}{\left(\frac{b}{\sqrt{C_b}}\right)}+377\left(\frac{b}{\sqrt{C_b}}\right)\sqrt{\left(\frac{b}{\sqrt{C_b}}\right)^2+1} \nonumber\\
		+&44\left(\frac{b}{\sqrt{C_b}}\right)^3\sqrt{\left(\frac{b}{\sqrt{C_b}}\right)^2+1}+135\sinh^{-1}{\left(\frac{b}{\sqrt{C_b}}\right)}\Bigg]\,,
		\label{eq:c2_GS} \\
		c_3^{\mathrm{GS}}=\frac{C_b^{5/2}}{80b^8}\Bigg[&80 \left(\frac{b}{\sqrt{C_b}}\right)^7\sqrt{\left(\frac{b}{\sqrt{C_b}}\right)^2+1}-1048 \left(\frac{b}{\sqrt{C_b}}\right)^5\sqrt{\left(\frac{b}{\sqrt{C_b}}\right)^2+1} \nonumber \\
		-&5766\left(\frac{b}{\sqrt{C_b}}\right)^3\sqrt{\left(\frac{b}{\sqrt{C_b}}\right)^2+1}+3584 \left(\frac{b}{\sqrt{C_b}}\right)^5 \nonumber\\
		- & 53883 \left(\frac{b}{\sqrt{C_b}}\right)\sqrt{\left(\frac{b}{\sqrt{C_b}}\right)^2+1}+73728\left(\frac{b}{\sqrt{C_b}}\right) \nonumber\\
        +&29400 \left(\frac{b}{\sqrt{C_b}}\right)^2\sinh^{-1}{\left(\frac{b}{\sqrt{C_b}}\right)}-19845\sinh^{-1}{\left(\frac{b}{\sqrt{C_b}}\right)}\Bigg]\,,
		\label{eq:c3_GS} \\
		c_4^{\mathrm{GS}} = \frac{63 C_b^{5/2}}{40 b^9}\Bigg[& 4 \left(\frac{b}{\sqrt{C_b}}\right)^5\sqrt{\left(\frac{b}{\sqrt{C_b}}\right)^2+1}-16\left(\frac{b}{\sqrt{C_b}}\right)^5+28\left(\frac{b}{\sqrt{C_b}}\right)^3\sqrt{\left(\frac{b}{\sqrt{C_b}}\right)^2+1}\nonumber \\
		-&150\left(\frac{b}{\sqrt{C_b}}\right)^2\sinh^{-1}{\left(\frac{b}{\sqrt{C_b}}\right)}+279\left(\frac{b}{\sqrt{C_b}}\right)\sqrt{\left(\frac{b}{\sqrt{C_b}}\right)^2+1} \nonumber\\
        -&384\left(\frac{b}{\sqrt{C_b}}\right) +105\sinh^{-1}{\left(\frac{b}{\sqrt{C_b}}\right)}\Bigg]\,.
		\label{eq:c4_GS}
\end{align}

In this case, we can directly write the Daisy contribution as a polynomial without introducing a tadpole term, given by
\begin{equation}
    V_{\mathrm{Daisy}}^{\mathrm{GS}}(\varphi,T)=V_{\mathrm{GS},0}+\frac{1}{2}m^2_{\mathrm{GS}}\varphi^2-\frac{1}{3}K_{\mathrm{GS}}\varphi^3-\frac{1}{4}\lambda_{\mathrm{GS}}\varphi^4\,,
    \label{eq:Daisy_GS}
\end{equation}
with the polynomial coefficients given by
\begin{equation}
    \begin{split}
        V_{\mathrm{GS},0}=&-\frac{C_b^{3/2}g'^3T^4}{12\pi}\,,\quad 
        m^2_{\mathrm{GS}}=-\frac{g'^3C_b^{3/2}T^2 c_2^{\mathrm{GS}}}{6\pi}\,,\\
        K_{\mathrm{GS}}=& \frac{g'^3 T}{4\pi}\left(C_b^{3/2} c_3^{\mathrm{GS}} -1\right)\,, \quad
        \lambda_{\mathrm{GS}}=  \frac{g'^3 C_b^{3/2} c_4^{\mathrm{GS}}}{3\pi}\,.
    \end{split}
    \label{eq:coefficients_Daisy_GS}
\end{equation}

\section{Non-minimal RSB}\label{app:running_non-minimal}
While we studied in detail the minimal case with RSB, it could be the case that there are other fields that couple to the scalar field undergoing the FOPT. Here we consider the case that some dark fermions also acquire their masses thanks to the generation of the scalar vev. We include a left and a right-handed dark fermion, $\nu_{D,L}$ and $\nu_{D,R}$ respectively, with opposite charges under the new $U(1)^{\prime}$ gauge symmetry. We take the charge of the scalar field to be 1, while the absolute value of the fermion fields charge is 1/2. The Lagrangian for such case is given by
\begin{equation}
    \begin{split}
        \mathcal{L}=&-\frac{1}{4}F'_{\mu\nu}F'^{\mu\nu}+\left(D_\mu\varphi\right)^{\dagger}\left(D^\mu\varphi\right)+i \bar{\nu}_{D,L}\slashed{D}\nu_{D,L}+i \bar{\nu}_{D,R}\slashed{D}\nu_{D,R}\\
        &- (\bar{\nu}_{D,L} y \nu_{D,R} \varphi + \mathrm{h.c.})-\frac{\lambda}{4}(\varphi^\dagger\varphi)^2\,.
    \end{split}
    \label{eq:lag_non-minimal}
\end{equation}
The RGE-flows one needs to take into account are
\begin{equation}
    \begin{cases}
        \frac{dg'}{dt}=&\frac{1}{16\pi^2}\frac{2 g'^3}{3}\,,\\
        \frac{d\lambda}{dt}=&\frac{1}{8\pi^2}\Big[10\lambda^2-6g'^2\lambda+3g'^4+2\lambda|y|^2-|y|^4\Big]\,,\\
        \frac{dy}{dt}=&\frac{y}{16\pi^2}\left(2|y|^2-\frac{3}{2}g'^2\right)\,,
    \end{cases}
\end{equation}
where we have left the dependence of the couplings on $t=\log{\mu/\mu_0}$, with $\mu_0=m_{Z',0}$, implicit to ease the notation. In contrast to the minimal scenario in Section~\ref{sec:RGE}, we now need to solve these equations numerically. Moreover, the scalar anomalous dimension now also receives further contributions, for which we find
\begin{equation}
    \gamma=\frac{1}{16\pi^2}\left[-3g'^2+|y|^2\right]\,.
\end{equation}

Regarding the polynomial approximations for $V_{\mathrm{eff}}$ in this case, we note that Daisy contributions are not affected by the presence of fermions, such that $V_{\mathrm{Daisy}}^{\mathrm{GS}\,(\mathrm{Leg})}$ from Eq.~(\ref{eq:Daisy_Poly}) can still be used. Instead, for $V_{\mathrm{HT}}$ in Eq.~(\ref{eq:polynomial_shape_pot}) the polynomial coefficients are given by
\begin{equation}
	\begin{split}
		&V_0 \equiv -\frac{13 \pi^2 T^4}{180}\,,\quad
		m^2(T) \equiv  \frac{T^2}{12}\left(3g'^2+|y|^2\right)\,,\quad
		K(T) \equiv \frac{3 T}{4\pi}g'^3\,,\\
		&\lambda(T) \equiv -\lambda -\frac{1}{16\pi^2}\left[3g'^4\left(\log{\frac{a_b T^2}{\mu^2}}-\frac{5}{6}\right)-|y|^4\left(\log{\frac{a_f T^2}{\mu^2}}-\frac{3}{2}\right)\right]\,.
	\end{split}
	\label{eq:polynomial_params_NoDaisy_nonmin}
\end{equation}

Finally, the fermion fields would also have an impact in the FOPT strength, which is in this case given by
\begin{equation}
    \alpha\simeq \frac{1}{\rho_{\mathrm{rad}}(T_p)}\frac{3g_0^{\prime4}-y_0^4}{128\pi^2}\varphi_0^4\,.
\end{equation}
\section{Tree-level SSB}\label{app:tree-level-SSB}
We comment here on the possibility to write down the effective potential as a polynomial for the case of a scalar potential with SSB at tree-level, as is the case in the SM. The tree-level potential for a single field leading to SSB is
\begin{equation}
    V_{\mathrm{tree}}(\varphi)=-\frac{1}{2}m_0^2 \varphi^2+\frac{1}{4}{\lambda}\varphi^4\,,
    \label{eq:Vtree_SSB}
\end{equation}
which has a minimum at $\varphi_0^2 \equiv \frac{m_0^2}{\lambda}$. In this case, we cannot neglect the scalar and Goldstone contributions to the effective potential. Their field-dependent masses are respectively given by
\begin{equation}
    m_{\varphi}^2 (\varphi) = \lambda\left(3\varphi^2-\varphi_0^2\right)\,,\quad m_G^2 (\varphi) =\lambda\left(\varphi^2-\varphi_0^2\right)\,,
    \label{eq:scalar_Goldstone_masses}
\end{equation}
while the scalar thermal mass is given by
\begin{equation}
    \Pi_{\varphi\,(G)}(T)=\frac{T^2}{12}\left(3g'^2+4\lambda\right)\,.
    \label{eq:thermal_mass_scalar}
\end{equation}

Neglecting again possible contributions from fermions, we can apply the HT expansion from Eq.~(\ref{eq:explicit_HT_expansion_pot}) to find the following shape of $V_{\mathrm{HT}}$:
\begin{align}
    V_{\mathrm{HT}}(\varphi,T)=&V_0+\frac{1}{2}\varphi^2\Bigg[-m_0^2+\frac{T^2}{12}\left(3g'^2+4\lambda\right)-\frac{\lambda}{4\pi^2}m_0^2\left(\log{\frac{a_bT^2}{\mu^2}}-\frac{3}{2}\right)\Bigg] \nonumber \\
    -& \frac{T}{4\pi}g'^3\varphi^3-\frac{T}{12\pi}\lambda^{3/2}\left[\left(3\varphi^2-\varphi_0^2\right)^{3/2}+\left(\varphi^2-\varphi_0^2\right)^{3/2}\right]\nonumber\\
    +&\frac{1}{4}\varphi^4\Bigg[\lambda+\frac{1}{16\pi^2}\left[3g'^4\left(\log{\frac{a_bT^2}{\mu^2}}-\frac{5}{6}\right)+10\lambda^2\left(\log{\frac{a_bT^2}{\mu^2}}-\frac{3}{2}\right)\right]\Bigg]\,.
    \label{eq:VHT_treelevel_SSB}
\end{align}
It is obvious here that the HT expansion does not automatically render the potential as a polynomial because of the structure of the scalar and Goldstone masses. We note two possibilities here:
\begin{itemize}
    \item Hierarchical spectrum: If there is a hierarchy between the gauge boson mass and the scalar masses, such that $g'^2\gg 2\lambda$, then the scalar contributions can be neglected altogether. This case then reduces to our discussion in the main text on RSB where, by construction, one finds $\lambda\sim g'^4\ll g'^2$ for perturbative couplings. In this case however this would only allow to study a particular region of parameter space. 
    \item Polynomial decomposition: One could go beyond the HT expansion and use the same approach we introduced for Daisy resummation by projecting the non-polynomial terms into a basis of orthogonal polynomials.
\end{itemize}

On top of this, one should additionally consider Daisy resummation in order to study the tunneling process, which in this case explicitly reads as
\begin{equation}
    \begin{split}
        V_{\mathrm{Daisy}}(\varphi,T)=&-\frac{T}{12\pi}\Bigg[\left(g'^2\varphi^2+\frac{g'^2}{3}T^2\right)^{3/2}-g'^3\varphi^3\\
        &+\left(3\lambda\varphi^2-\lambda\varphi_0^2+\frac{3g'^2+4\lambda}{12}T^2\right)^{3/2}-\lambda^{3/2}\left(3\varphi^2-\varphi_0^2\right)^{3/2}\\
        &+\left(\lambda\varphi^2-\lambda\varphi_0^2+\frac{3g'^2+4\lambda}{12}T^2\right)^{3/2}-\lambda^{3/2}\left(\varphi^2-\varphi_0^2\right)^{3/2}\Bigg]\,.
    \end{split}
    \label{eq:Daisy_tree_SSB}
\end{equation}
We consider that performing the polynomial decomposition in this case would need to introduce different scales $b$ (see the discussion in Subsection~\ref{sec:Daisy}) for each term, considerably degrading the approximation. More importantly, in this case the potential barrier does not necessarily appear at small field values $\varphi\ll T$, such that the HT approximation is not as justified as in the classically scale-invariant scenario we studied in the main text. For these reasons, we consider that scenarios with SSB at tree-level cannot be consistently studied as a polynomial potential with our current approximations, for which we can use the results of Ref.~\cite{Levi:2022bzt} to find $S_3$ reliably. 

\bibliographystyle{JHEP} 
\bibliography{bibliography} 

@article{NANOGrav:2023hde,
    author = "Agazie, Gabriella and others",
    collaboration = "NANOGrav",
    title = "{The NANOGrav 15 yr Data Set: Observations and Timing of 68 Millisecond Pulsars}",
    eprint = "2306.16217",
    archivePrefix = "arXiv",
    primaryClass = "astro-ph.HE",
    doi = "10.3847/2041-8213/acda9a",
    journal = "Astrophys. J. Lett.",
    volume = "951",
    number = "1",
    pages = "L9",
    year = "2023"
}

@article{Kierkla:2023von,
    author = "Kierkla, Maciej and Swiezewska, Bogumila and Tenkanen, Tuomas V. I. and van de Vis, Jorinde",
    title = "{Gravitational waves from supercooled phase transitions: dimensional transmutation meets dimensional reduction}",
    eprint = "2312.12413",
    archivePrefix = "arXiv",
    primaryClass = "hep-ph",
    doi = "10.1007/JHEP02(2024)234",
    journal = "JHEP",
    volume = "02",
    pages = "234",
    year = "2024"
}

@article{Gould:2021ccf,
    author = "Gould, Oliver and Hirvonen, Joonas",
    title = "{Effective field theory approach to thermal bubble nucleation}",
    eprint = "2108.04377",
    archivePrefix = "arXiv",
    primaryClass = "hep-ph",
    reportNumber = "HIP-2020-19/TH",
    doi = "10.1103/PhysRevD.104.096015",
    journal = "Phys. Rev. D",
    volume = "104",
    number = "9",
    pages = "096015",
    year = "2021"
}

@article{Kierkla:2025qyz,
    author = "Kierkla, Maciej and Schicho, Philipp and Swiezewska, Bogumila and Tenkanen, Tuomas V. I. and van de Vis, Jorinde",
    title = "{Finite-temperature bubble nucleation with shifting scale hierarchies}",
    eprint = "2503.13597",
    archivePrefix = "arXiv",
    primaryClass = "hep-ph",
    reportNumber = "CERN-TH-2025-046, HIP-2024-27/TH",
    doi = "10.1007/JHEP07(2025)153",
    journal = "JHEP",
    volume = "07",
    pages = "153",
    year = "2025"
}

@article{Kierkla:2025vwp,
    author = "Kierkla, Maciej and Ramberg, Nicklas and Schicho, Philipp and Schmitt, Daniel",
    title = "{Theoretical uncertainties for primordial black holes from cosmological phase transitions}",
    eprint = "2506.15496",
    archivePrefix = "arXiv",
    primaryClass = "hep-ph",
    reportNumber = "SISSA 07/2025/FISI",
    month = "6",
    year = "2025"
}

@article{lamb2023rapid,
  title={Rapid refitting techniques for Bayesian spectral characterization of the gravitational wave background using pulsar timing arrays},
  author={Lamb, William G and Taylor, Stephen R and van Haasteren, Rutger},
  journal={Physical Review D},
  volume={108},
  number={10},
  pages={103019},
  year={2023},
  publisher={APS}
}

@article{Hirvonen:2024rfg,
    author = "Hirvonen, Joonas",
    title = "{Real-time nucleation and off-equilibrium effects in high-temperature quantum field theories}",
    eprint = "2403.07987",
    archivePrefix = "arXiv",
    primaryClass = "hep-ph",
    reportNumber = "HIP-2024-6/TH",
    doi = "10.1103/4p7n-nyln",
    journal = "Phys. Rev. D",
    volume = "111",
    number = "11",
    pages = "116020",
    year = "2025"
}

@article{Espinosa:2018szu,
    author = "Espinosa, J. R. and Konstandin, T.",
    title = "{A Fresh Look at the Calculation of Tunneling Actions in Multi-Field Potentials}",
    eprint = "1811.09185",
    archivePrefix = "arXiv",
    primaryClass = "hep-th",
    reportNumber = "DESY-18-206, CERN-TH-2018-245",
    doi = "10.1088/1475-7516/2019/01/051",
    journal = "JCAP",
    volume = "01",
    pages = "051",
    year = "2019"
}

@article{Mitridate:2023oar,
    author = {Mitridate, Andrea and Wright, David and von Eckardstein, Richard and Schr{\"o}der, Tobias and Nay, Jonathan and Olum, Ken and Schmitz, Kai and Trickle, Tanner},
    title = "{PTArcade}",
    eprint = "2306.16377",
    archivePrefix = "arXiv",
    primaryClass = "hep-ph",
    reportNumber = "FERMILAB-PUB-23-588-T",
    month = "6",
    year = "2023"
}

@article{Wainwright:2011kj,
    author = "Wainwright, Carroll L.",
    title = "{CosmoTransitions: Computing Cosmological Phase Transition Temperatures and Bubble Profiles with Multiple Fields}",
    eprint = "1109.4189",
    archivePrefix = "arXiv",
    primaryClass = "hep-ph",
    doi = "10.1016/j.cpc.2012.04.004",
    journal = "Comput. Phys. Commun.",
    volume = "183",
    pages = "2006--2013",
    year = "2012"
}

@article{Gouttenoire:2021kjv,
    author = "Gouttenoire, Yann and Jinno, Ryusuke and Sala, Filippo",
    title = "{Friction pressure on relativistic bubble walls}",
    eprint = "2112.07686",
    archivePrefix = "arXiv",
    primaryClass = "hep-ph",
    reportNumber = "DESY-21-147, IFT-UAM/CSIC-21-146",
    doi = "10.1007/JHEP05(2022)004",
    journal = "JHEP",
    volume = "05",
    pages = "004",
    year = "2022"
}

@article{Bodeker:2017cim,
    author = "Bodeker, Dietrich and Moore, Guy D.",
    title = "{Electroweak Bubble Wall Speed Limit}",
    eprint = "1703.08215",
    archivePrefix = "arXiv",
    primaryClass = "hep-ph",
    doi = "10.1088/1475-7516/2017/05/025",
    journal = "JCAP",
    volume = "05",
    pages = "025",
    year = "2017"
}

@article{Bodeker:2009qy,
    author = "Bodeker, Dietrich and Moore, Guy D.",
    title = "{Can electroweak bubble walls run away?}",
    eprint = "0903.4099",
    archivePrefix = "arXiv",
    primaryClass = "hep-ph",
    doi = "10.1088/1475-7516/2009/05/009",
    journal = "JCAP",
    volume = "05",
    pages = "009",
    year = "2009"
}

@article{Ellis:2019oqb,
    author = "Ellis, John and Lewicki, Marek and No, Jos\'e Miguel and Vaskonen, Ville",
    title = "{Gravitational wave energy budget in strongly supercooled phase transitions}",
    eprint = "1903.09642",
    archivePrefix = "arXiv",
    primaryClass = "hep-ph",
    reportNumber = "KCL-PH-TH/2019-32, CERN-TH-2019-032, IFT-UAM/CSIC-19-32",
    doi = "10.1088/1475-7516/2019/06/024",
    journal = "JCAP",
    volume = "06",
    pages = "024",
    year = "2019"
}

@article{10.1063/1.1338506,
    author = {Lorenz, Christian D. and Ziff, Robert M.},
    title = {Precise determination of the critical percolation threshold for the three-dimensional “Swiss cheese” model using a growth algorithm},
    journal = {The Journal of Chemical Physics},
    volume = {114},
    number = {8},
    pages = {3659-3661},
    year = {2001},
    month = {02},
    issn = {0021-9606},
    doi = {10.1063/1.1338506},
    url = {https://doi.org/10.1063/1.1338506},
    eprint = {https://pubs.aip.org/aip/jcp/article-pdf/114/8/3659/19017114/3659\_1\_online.pdf},
}

@article{Athron:2023xlk,
    author = "Athron, Peter and Bal\'azs, Csaba and Fowlie, Andrew and Morris, Lachlan and Wu, Lei",
    title = "{Cosmological phase transitions: From perturbative particle physics to gravitational waves}",
    eprint = "2305.02357",
    archivePrefix = "arXiv",
    primaryClass = "hep-ph",
    doi = "10.1016/j.ppnp.2023.104094",
    journal = "Prog. Part. Nucl. Phys.",
    volume = "135",
    pages = "104094",
    year = "2024"
}

@article{Linde:1980tt,
    author = "Linde, Andrei D.",
    title = "{Fate of the False Vacuum at Finite Temperature: Theory and Applications}",
    reportNumber = "LEBEDEV-80-92",
    doi = "10.1016/0370-2693(81)90281-1",
    journal = "Phys. Lett. B",
    volume = "100",
    pages = "37--40",
    year = "1981"
}

@article{Coleman:1973jx,
    author = "Coleman, Sidney R. and Weinberg, Erick J.",
    title = "{Radiative Corrections as the Origin of Spontaneous Symmetry Breaking}",
    doi = "10.1103/PhysRevD.7.1888",
    journal = "Phys. Rev. D",
    volume = "7",
    pages = "1888--1910",
    year = "1973"
}

@article{Costa:2025csj,
    author = "Costa, Francesco and Hoefken Zink, Jaime and Lucente, Michele and Pascoli, Silvia and Rosauro-Alcaraz, Salvador",
    title = "{Supercooled dark scalar phase transitions explanation of NANOGrav data}",
    eprint = "2501.15649",
    archivePrefix = "arXiv",
    primaryClass = "hep-ph",
    doi = "10.1016/j.physletb.2025.139634",
    journal = "Phys. Lett. B",
    volume = "868",
    pages = "139634",
    year = "2025"
}

@article{Coleman:1977py,
    author = "Coleman, Sidney R.",
    title = "{The Fate of the False Vacuum. 1. Semiclassical Theory}",
    reportNumber = "HUTP-77-A004",
    doi = "10.1103/PhysRevD.16.1248",
    journal = "Phys. Rev. D",
    volume = "15",
    pages = "2929--2936",
    year = "1977",
    note = "[Erratum: Phys.Rev.D 16, 1248 (1977)]"
}

@article{Espinosa:2018hue,
    author = "Espinosa, J. R.",
    title = "{A Fresh Look at the Calculation of Tunneling Actions}",
    eprint = "1805.03680",
    archivePrefix = "arXiv",
    primaryClass = "hep-th",
    doi = "10.1088/1475-7516/2018/07/036",
    journal = "JCAP",
    volume = "07",
    pages = "036",
    year = "2018"
}

@article{NANOGrav:2023gor,
    author = "Agazie, Gabriella and others",
    collaboration = "NANOGrav",
    title = "{The NANOGrav 15 yr Data Set: Evidence for a Gravitational-wave Background}",
    eprint = "2306.16213",
    archivePrefix = "arXiv",
    primaryClass = "astro-ph.HE",
    doi = "10.3847/2041-8213/acdac6",
    journal = "Astrophys. J. Lett.",
    volume = "951",
    number = "1",
    pages = "L8",
    year = "2023"
}

@article{NANOGrav:2023pdq,
    author = "Agazie, Gabriella and others",
    collaboration = "NANOGrav",
    title = "{The NANOGrav 15 yr Data Set: Bayesian Limits on Gravitational Waves from Individual Supermassive Black Hole Binaries}",
    eprint = "2306.16222",
    archivePrefix = "arXiv",
    primaryClass = "astro-ph.HE",
    doi = "10.3847/2041-8213/ace18a",
    journal = "Astrophys. J. Lett.",
    volume = "951",
    number = "2",
    pages = "L50",
    year = "2023"
}

@article{NANOGrav:2023hfp,
    author = "Agazie, Gabriella and others",
    collaboration = "NANOGrav",
    title = "{The NANOGrav 15 yr Data Set: Constraints on Supermassive Black Hole Binaries from the Gravitational-wave Background}",
    eprint = "2306.16220",
    archivePrefix = "arXiv",
    primaryClass = "astro-ph.HE",
    doi = "10.3847/2041-8213/ace18b",
    journal = "Astrophys. J. Lett.",
    volume = "952",
    number = "2",
    pages = "L37",
    year = "2023"
}

@article{EPTA:2023fyk,
    author = "Antoniadis, J. and others",
    collaboration = "EPTA, InPTA:",
    title = "{The second data release from the European Pulsar Timing Array - III. Search for gravitational wave signals}",
    eprint = "2306.16214",
    archivePrefix = "arXiv",
    primaryClass = "astro-ph.HE",
    doi = "10.1051/0004-6361/202346844",
    journal = "Astron. Astrophys.",
    volume = "678",
    pages = "A50",
    year = "2023"
}

@article{Reardon:2023gzh,
    author = "Reardon, Daniel J. and others",
    title = "{Search for an Isotropic Gravitational-wave Background with the Parkes Pulsar Timing Array}",
    eprint = "2306.16215",
    archivePrefix = "arXiv",
    primaryClass = "astro-ph.HE",
    doi = "10.3847/2041-8213/acdd02",
    journal = "Astrophys. J. Lett.",
    volume = "951",
    number = "1",
    pages = "L6",
    year = "2023"
}

@article{Xu:2023wog,
    author = "Xu, Heng and others",
    title = "{Searching for the Nano-Hertz Stochastic Gravitational Wave Background with the Chinese Pulsar Timing Array Data Release I}",
    eprint = "2306.16216",
    archivePrefix = "arXiv",
    primaryClass = "astro-ph.HE",
    doi = "10.1088/1674-4527/acdfa5",
    journal = "Res. Astron. Astrophys.",
    volume = "23",
    number = "7",
    pages = "075024",
    year = "2023"
}

@article{Wyithe_2003,
doi = {10.1086/375187},
url = {https://dx.doi.org/10.1086/375187},
year = {2003},
month = {jun},
publisher = {},
volume = {590},
number = {2},
pages = {691},
author = {Wyithe, J. Stuart B. and Loeb, Abraham},
title = {Low-Frequency Gravitational Waves from Massive Black Hole Binaries: Predictions for LISA and Pulsar Timing Arrays},
journal = {The Astrophysical Journal}
}

@article{Jaffe_2003,
doi = {10.1086/345443},
url = {https://dx.doi.org/10.1086/345443},
year = {2003},
month = {feb},
publisher = {},
volume = {583},
number = {2},
pages = {616},
author = {Jaffe, A. H. and Backer, D. C.},
title = {Gravitational Waves Probe the Coalescence Rate of Massive Black Hole Binaries},
journal = {The Astrophysical Journal}
}

@article{Gouttenoire:2023bqy,
    author = "Gouttenoire, Yann",
    title = "{First-Order Phase Transition Interpretation of Pulsar Timing Array Signal Is Consistent with Solar-Mass Black Holes}",
    eprint = "2307.04239",
    archivePrefix = "arXiv",
    primaryClass = "hep-ph",
    doi = "10.1103/PhysRevLett.131.171404",
    journal = "Phys. Rev. Lett.",
    volume = "131",
    number = "17",
    pages = "171404",
    year = "2023"
}

@article{Chen:2023bms,
    author = "Chen, Zu-Cheng and Li, Shou-Long and Wu, Puxun and Yu, Hongwei",
    title = "{NANOGrav hints for first-order confinement-deconfinement phase transition in different QCD-matter scenarios}",
    eprint = "2312.01824",
    archivePrefix = "arXiv",
    primaryClass = "astro-ph.CO",
    doi = "10.1103/PhysRevD.109.043022",
    journal = "Phys. Rev. D",
    volume = "109",
    number = "4",
    pages = "043022",
    year = "2024"
}

@article{Wang:2023bbc,
    author = "Wang, Deng",
    title = "{Constraining Cosmological Phase Transitions with Chinese Pulsar Timing Array Data Release 1}",
    eprint = "2307.15970",
    archivePrefix = "arXiv",
    primaryClass = "astro-ph.CO",
    month = "7",
    year = "2023"
}

@article{Ghosh:2023aum,
    author = "Ghosh, Tathagata and Ghoshal, Anish and Guo, Huai-Ke and Hajkarim, Fazlollah and King, Stephen F. and Sinha, Kuver and Wang, Xin and White, Graham",
    title = "{Did we hear the sound of the Universe boiling? Analysis using the full fluid velocity profiles and NANOGrav 15-year data}",
    eprint = "2307.02259",
    archivePrefix = "arXiv",
    primaryClass = "astro-ph.HE",
    doi = "10.1088/1475-7516/2024/05/100",
    journal = "JCAP",
    volume = "05",
    pages = "100",
    year = "2024"
}

@article{Addazi:2023jvg,
    author = "Addazi, Andrea and Cai, Yi-Fu and Marciano, Antonino and Visinelli, Luca",
    title = "{Have pulsar timing array methods detected a cosmological phase transition?}",
    eprint = "2306.17205",
    archivePrefix = "arXiv",
    primaryClass = "astro-ph.CO",
    reportNumber = "CA21106; CA21136",
    doi = "10.1103/PhysRevD.109.015028",
    journal = "Phys. Rev. D",
    volume = "109",
    number = "1",
    pages = "015028",
    year = "2024"
}

@article{Bringmann:2023opz,
    author = "Bringmann, Torsten and Depta, Paul Frederik and Konstandin, Thomas and Schmidt-Hoberg, Kai and Tasillo, Carlo",
    title = "{Does NANOGrav observe a dark sector phase transition?}",
    eprint = "2306.09411",
    archivePrefix = "arXiv",
    primaryClass = "astro-ph.CO",
    reportNumber = "DESY-23-077",
    doi = "10.1088/1475-7516/2023/11/053",
    journal = "JCAP",
    volume = "11",
    pages = "053",
    year = "2023"
}

@article{Croon:2024mde,
    author = "Croon, Djuna and Weir, David J.",
    title = "{Gravitational Waves from Phase Transitions}",
    eprint = "2410.21509",
    archivePrefix = "arXiv",
    primaryClass = "hep-ph",
    reportNumber = "HIP-2024-23/TH",
    month = "10",
    year = "2024"
}

@article{Winkler:2024olr,
    author = "Winkler, Martin Wolfgang and Freese, Katherine",
    title = "{Origin of the Stochastic Gravitational Wave Background: First-Order Phase Transition vs. Black Hole Mergers}",
    eprint = "2401.13729",
    archivePrefix = "arXiv",
    primaryClass = "astro-ph.CO",
    reportNumber = "NORDITA-2024-002, UT-WI-02-2024",
    month = "1",
    year = "2024"
}

@article{Conaci:2024tlc,
    author = "Conaci, Angela and Delle Rose, Luigi and Dev, P. S. Bhupal and Ghoshal, Anish",
    title = "{Slaying axion-like particles via gravitational waves and primordial black holes from supercooled phase transition}",
    eprint = "2401.09411",
    archivePrefix = "arXiv",
    primaryClass = "astro-ph.CO",
    doi = "10.1007/JHEP12(2024)196",
    journal = "JHEP",
    volume = "12",
    pages = "196",
    year = "2024"
}

@article{Ekstedt:2022bff,
    author = "Ekstedt, Andreas and Schicho, Philipp and Tenkanen, Tuomas V. I.",
    title = "{DRalgo: A package for effective field theory approach for thermal phase transitions}",
    eprint = "2205.08815",
    archivePrefix = "arXiv",
    primaryClass = "hep-ph",
    reportNumber = "HIP-2022-11/TH, NORDITA 2022-030",
    doi = "10.1016/j.cpc.2023.108725",
    journal = "Comput. Phys. Commun.",
    volume = "288",
    pages = "108725",
    year = "2023"
}

@article{Goncalves:2025uwh,
    author = "Gon{\c{c}}alves, Jo{\~a}o and Marfatia, Danny and Morais, Ant{\'o}nio P. and Pasechnik, Roman",
    title = "{Supercooled phase transitions in conformal dark sectors explain NANOGrav data}",
    eprint = "2501.11619",
    archivePrefix = "arXiv",
    primaryClass = "hep-ph",
    doi = "10.1016/j.physletb.2025.139829",
    journal = "Phys. Lett. B",
    volume = "869",
    pages = "139829",
    year = "2025"
}

@article{PhysRevD.7.1888,
  title = {Radiative Corrections as the Origin of Spontaneous Symmetry Breaking},
  author = {Coleman, Sidney and Weinberg, Erick},
  journal = {Phys. Rev. D},
  volume = {7},
  issue = {6},
  pages = {1888--1910},
  numpages = {0},
  year = {1973},
  month = {Mar},
  publisher = {American Physical Society},
  doi = {10.1103/PhysRevD.7.1888},
  url = {https://link.aps.org/doi/10.1103/PhysRevD.7.1888}
}

@article{PhysRevD.9.3320,
  title = {Symmetry behavior at finite temperature},
  author = {Dolan, L. and Jackiw, R.},
  journal = {Phys. Rev. D},
  volume = {9},
  issue = {12},
  pages = {3320--3341},
  numpages = {0},
  year = {1974},
  month = {Jun},
  publisher = {American Physical Society},
  doi = {10.1103/PhysRevD.9.3320},
  url = {https://link.aps.org/doi/10.1103/PhysRevD.9.3320}
}

@article{PhysRevD.45.2933,
  title = {Effective potential at finite temperature in the standard model},
  author = {Carrington, M. E.},
  journal = {Phys. Rev. D},
  volume = {45},
  issue = {8},
  pages = {2933--2944},
  numpages = {0},
  year = {1992},
  month = {Apr},
  publisher = {American Physical Society},
  doi = {10.1103/PhysRevD.45.2933},
  url = {https://link.aps.org/doi/10.1103/PhysRevD.45.2933}
}

@inproceedings{Quiros:1999jp,
    author = "Quiros, Mariano",
    title = "{Finite temperature field theory and phase transitions}",
    booktitle = "{ICTP Summer School in High-Energy Physics and Cosmology}",
    eprint = "hep-ph/9901312",
    archivePrefix = "arXiv",
    reportNumber = "IEM-FT-187-99",
    pages = "187--259",
    month = "1",
    year = "1999"
}

@article{Salvio:2023qgb,
    author = "Salvio, Alberto",
    title = "{Model-independent radiative symmetry breaking and gravitational waves}",
    eprint = "2302.10212",
    archivePrefix = "arXiv",
    primaryClass = "hep-ph",
    doi = "10.1088/1475-7516/2023/04/051",
    journal = "JCAP",
    volume = "04",
    pages = "051",
    year = "2023"
}

@article{Salvio:2023ynn,
    author = "Salvio, Alberto",
    title = "{Supercooling in radiative symmetry breaking: theory extensions, gravitational wave detection and primordial black holes}",
    eprint = "2307.04694",
    archivePrefix = "arXiv",
    primaryClass = "hep-ph",
    doi = "10.1088/1475-7516/2023/12/046",
    journal = "JCAP",
    volume = "12",
    pages = "046",
    year = "2023"
}

@article{Salvio:2023blb,
    author = "Salvio, Alberto",
    title = "{Pulsar timing arrays and primordial black holes from a supercooled phase transition}",
    eprint = "2312.04628",
    archivePrefix = "arXiv",
    primaryClass = "hep-ph",
    doi = "10.1016/j.physletb.2024.138639",
    journal = "Phys. Lett. B",
    volume = "852",
    pages = "138639",
    year = "2024"
}

@article{Levi:2022bzt,
    author = "Levi, Noam and Opferkuch, Toby and Redigolo, Diego",
    title = "{The supercooling window at weak and strong coupling}",
    eprint = "2212.08085",
    archivePrefix = "arXiv",
    primaryClass = "hep-ph",
    doi = "10.1007/JHEP02(2023)125",
    journal = "JHEP",
    volume = "02",
    pages = "125",
    year = "2023"
}

@article{Balan:2025uke,
    author = "Balan, Sowmiya and Bringmann, Torsten and Kahlhoefer, Felix and Matuszak, Jonas and Tasillo, Carlo",
    title = "{Sub-GeV dark matter and nano-Hertz gravitational waves from a classically conformal dark sector}",
    eprint = "2502.19478",
    archivePrefix = "arXiv",
    primaryClass = "hep-ph",
    doi = "10.1088/1475-7516/2025/08/062",
    journal = "JCAP",
    volume = "08",
    pages = "062",
    year = "2025"
}

@article{Arnold:1992rz,
    author = "Arnold, Peter Brockway and Espinosa, Olivier",
    title = "{The Effective potential and first order phase transitions: Beyond leading-order}",
    eprint = "hep-ph/9212235",
    archivePrefix = "arXiv",
    reportNumber = "UW-PT-92-18, USM-TH-60",
    doi = "10.1103/PhysRevD.47.3546",
    journal = "Phys. Rev. D",
    volume = "47",
    pages = "3546",
    year = "1993",
    note = "[Erratum: Phys.Rev.D 50, 6662 (1994)]"
}

@article{Christiansen:2025xhv,
    author = "Christiansen, Martin and Madge, Eric and Puchades-Ib{\'a}{\~n}ez, Cristina and Ramirez-Quezada, Maura E. and Schwaller, Pedro",
    title = "{Beyond the Daisy Chain: Running and the 3D EFT View of Supercooled Phase Transitions}",
    eprint = "2511.02910",
    archivePrefix = "arXiv",
    primaryClass = "hep-ph",
    reportNumber = "IFT-UAM/CSIC 25-139 and MITP-25-071",
    month = "11",
    year = "2025"
}

@article{Meissner:2008uw,
    author = "Meissner, Krzysztof A. and Nicolai, Hermann",
    title = "{Renormalization Group and Effective Potential in Classically Conformal Theories}",
    eprint = "0809.1338",
    archivePrefix = "arXiv",
    primaryClass = "hep-th",
    reportNumber = "AEI-2008-062, LPTENS-08-50",
    journal = "Acta Phys. Polon. B",
    volume = "40",
    pages = "2737--2752",
    year = "2009"
}

@article{Chataignier:2018kay,
    author = "Chataignier, Leonardo and Prokopec, Tomislav and Schmidt, Michael G. and {\'S}wie{\.z}ewska, Bogumi{\l}a",
    title = "{Systematic analysis of radiative symmetry breaking in models with extended scalar sector}",
    eprint = "1805.09292",
    archivePrefix = "arXiv",
    primaryClass = "hep-ph",
    doi = "10.1007/JHEP08(2018)083",
    journal = "JHEP",
    volume = "08",
    pages = "083",
    year = "2018"
}

@article{Chataignier:2018aud,
    author = "Chataignier, Leonardo and Prokopec, Tomislav and Schmidt, Michael G. and Swiezewska, Bogumila",
    title = "{Single-scale Renormalisation Group Improvement of Multi-scale Effective Potentials}",
    eprint = "1801.05258",
    archivePrefix = "arXiv",
    primaryClass = "hep-ph",
    doi = "10.1007/JHEP03(2018)014",
    journal = "JHEP",
    volume = "03",
    pages = "014",
    year = "2018"
}

@article{Ellis:2020nnr,
    author = "Ellis, John and Lewicki, Marek and Vaskonen, Ville",
    title = "{Updated predictions for gravitational waves produced in a strongly supercooled phase transition}",
    eprint = "2007.15586",
    archivePrefix = "arXiv",
    primaryClass = "astro-ph.CO",
    reportNumber = "KCL-PH-TH/2020-40, CERN-TH-2020-129",
    doi = "10.1088/1475-7516/2020/11/020",
    journal = "JCAP",
    volume = "11",
    pages = "020",
    year = "2020"
}

@article{Lewicki:2024xan,
    author = "Lewicki, Marek and Merchand, Marco and Sagunski, Laura and Schicho, Philipp and Schmitt, Daniel",
    title = "{Impact of theoretical uncertainties on model parameter reconstruction from GW signals sourced by cosmological phase transitions}",
    eprint = "2403.03769",
    archivePrefix = "arXiv",
    primaryClass = "hep-ph",
    doi = "10.1103/PhysRevD.110.023538",
    journal = "Phys. Rev. D",
    volume = "110",
    number = "2",
    pages = "023538",
    year = "2024"
}

@article{Gould:2021oba,
    author = "Gould, Oliver and Tenkanen, Tuomas V. I.",
    title = "{On the perturbative expansion at high temperature and implications for cosmological phase transitions}",
    eprint = "2104.04399",
    archivePrefix = "arXiv",
    primaryClass = "hep-ph",
    reportNumber = "NORDITA 2021-010",
    doi = "10.1007/JHEP06(2021)069",
    journal = "JHEP",
    volume = "06",
    pages = "069",
    year = "2021"
}

@article{Croon:2020cgk,
    author = "Croon, Djuna and Gould, Oliver and Schicho, Philipp and Tenkanen, Tuomas V. I. and White, Graham",
    title = "{Theoretical uncertainties for cosmological first-order phase transitions}",
    eprint = "2009.10080",
    archivePrefix = "arXiv",
    primaryClass = "hep-ph",
    reportNumber = "HIP-2020-26/TH",
    doi = "10.1007/JHEP04(2021)055",
    journal = "JHEP",
    volume = "04",
    pages = "055",
    year = "2021"
}

@article{Athron:2023rfq,
    author = "Athron, Peter and Morris, Lachlan and Xu, Zhongxiu",
    title = "{How robust are gravitational wave predictions from cosmological phase transitions?}",
    eprint = "2309.05474",
    archivePrefix = "arXiv",
    primaryClass = "hep-ph",
    doi = "10.1088/1475-7516/2024/05/075",
    journal = "JCAP",
    volume = "05",
    pages = "075",
    year = "2024"
}

@article{Poole:2019kcm,
    author = "Poole, Colin and Thomsen, Anders Eller",
    title = "{Constraints on 3- and 4-loop $\beta$-functions in a general four-dimensional Quantum Field Theory}",
    eprint = "1906.04625",
    archivePrefix = "arXiv",
    primaryClass = "hep-th",
    doi = "10.1007/JHEP09(2019)055",
    journal = "JHEP",
    volume = "09",
    pages = "055",
    year = "2019"
}

@article{Xu:2025zsv,
    author = "Xu, Rui and Lu, Jiachen and Deng, Shihao and Bian, Ligong",
    title = "{Phase Transitions in the Early Universe: Impacts on BBN and CMB Observables}",
    eprint = "2503.19737",
    archivePrefix = "arXiv",
    primaryClass = "hep-ph",
    month = "3",
    year = "2025"
}

@article{Bai:2021ibt,
    author = "Bai, Yang and Korwar, Mrunal",
    title = "{Cosmological constraints on first-order phase transitions}",
    eprint = "2109.14765",
    archivePrefix = "arXiv",
    primaryClass = "hep-ph",
    doi = "10.1103/PhysRevD.105.095015",
    journal = "Phys. Rev. D",
    volume = "105",
    number = "9",
    pages = "095015",
    year = "2022"
}

@article{Rajagopal:1994zj,
    author = "Rajagopal, Mohan and Romani, Roger W.",
    title = "{Ultralow frequency gravitational radiation from massive black hole binaries}",
    eprint = "astro-ph/9412038",
    archivePrefix = "arXiv",
    doi = "10.1086/175813",
    journal = "Astrophys. J.",
    volume = "446",
    pages = "543--549",
    year = "1995"
}

@article{Ellis:2023oxs,
    author = {Ellis, John and Fairbairn, Malcolm and Franciolini, Gabriele and H{\"u}tsi, Gert and Iovino, Antonio and Lewicki, Marek and Raidal, Martti and Urrutia, Juan and Vaskonen, Ville and Veerm{\"a}e, Hardi},
    title = "{What is the source of the PTA GW signal?}",
    eprint = "2308.08546",
    archivePrefix = "arXiv",
    primaryClass = "astro-ph.CO",
    reportNumber = "KCL-PH-TH/2023-43, CERN-TH-2023-153, AION-REPORT/2023-08",
    doi = "10.1103/PhysRevD.109.023522",
    journal = "Phys. Rev. D",
    volume = "109",
    number = "2",
    pages = "023522",
    year = "2024"
}

@article{Figueroa:2023zhu,
    author = "Figueroa, Daniel G. and Pieroni, Mauro and Ricciardone, Angelo and Simakachorn, Peera",
    title = "{Cosmological Background Interpretation of Pulsar Timing Array Data}",
    eprint = "2307.02399",
    archivePrefix = "arXiv",
    primaryClass = "astro-ph.CO",
    reportNumber = "CERN-TH-2023-132",
    doi = "10.1103/PhysRevLett.132.171002",
    journal = "Phys. Rev. Lett.",
    volume = "132",
    number = "17",
    pages = "171002",
    year = "2024"
}

@article{Madge:2023dxc,
    author = "Madge, Eric and Morgante, Enrico and Puchades-Ib{\'a}{\~n}ez, Cristina and Ramberg, Nicklas and Ratzinger, Wolfram and Schenk, Sebastian and Schwaller, Pedro",
    title = "{Primordial gravitational waves in the nano-Hertz regime and PTA data {\textemdash} towards solving the GW inverse problem}",
    eprint = "2306.14856",
    archivePrefix = "arXiv",
    primaryClass = "hep-ph",
    reportNumber = "MITP-23-029",
    doi = "10.1007/JHEP10(2023)171",
    journal = "JHEP",
    volume = "10",
    pages = "171",
    year = "2023"
}

@article{Wu:2023hsa,
    author = "Wu, Yu-Mei and Chen, Zu-Cheng and Huang, Qing-Guo",
    title = "{Cosmological interpretation for the stochastic signal in pulsar timing arrays}",
    eprint = "2307.03141",
    archivePrefix = "arXiv",
    primaryClass = "astro-ph.CO",
    doi = "10.1007/s11433-023-2298-7",
    journal = "Sci. China Phys. Mech. Astron.",
    volume = "67",
    number = "4",
    pages = "240412",
    year = "2024"
}

@article{Caprini:2024hue,
    author = "Caprini, Chiara and Jinno, Ryusuke and Lewicki, Marek and Madge, Eric and Merchand, Marco and Nardini, Germano and Pieroni, Mauro and Roper Pol, Alberto and Vaskonen, Ville",
    collaboration = "LISA Cosmology Working Group",
    title = "{Gravitational waves from first-order phase transitions in LISA: reconstruction pipeline and physics interpretation}",
    eprint = "2403.03723",
    archivePrefix = "arXiv",
    primaryClass = "astro-ph.CO",
    reportNumber = "LISA-COSWG-24-01, CERN-TH-2024-029",
    doi = "10.1088/1475-7516/2024/10/020",
    journal = "JCAP",
    volume = "10",
    pages = "020",
    year = "2024"
}

@article{PhysRevD.30.272,
  title = {Cosmic separation of phases},
  author = {Witten, Edward},
  journal = {Phys. Rev. D},
  volume = {30},
  issue = {2},
  pages = {272--285},
  numpages = {0},
  year = {1984},
  month = {Jul},
  publisher = {American Physical Society},
  doi = {10.1103/PhysRevD.30.272},
  url = {https://link.aps.org/doi/10.1103/PhysRevD.30.272}
}

@article{Athron:2023mer,
    author = "Athron, Peter and Fowlie, Andrew and Lu, Chih-Ting and Morris, Lachlan and Wu, Lei and Wu, Yongcheng and Xu, Zhongxiu",
    title = "{Can Supercooled Phase Transitions Explain the Gravitational Wave Background Observed by Pulsar Timing Arrays?}",
    eprint = "2306.17239",
    archivePrefix = "arXiv",
    primaryClass = "hep-ph",
    doi = "10.1103/PhysRevLett.132.221001",
    journal = "Phys. Rev. Lett.",
    volume = "132",
    number = "22",
    pages = "221001",
    year = "2024"
}

@article{NANOGrav:2023hvm,
    author = "Afzal, Adeela and others",
    collaboration = "NANOGrav",
    title = "{The NANOGrav 15 yr Data Set: Search for Signals from New Physics}",
    eprint = "2306.16219",
    archivePrefix = "arXiv",
    primaryClass = "astro-ph.HE",
    reportNumber = "FERMILAB-PUB-23-589-T",
    doi = "10.3847/2041-8213/acdc91",
    journal = "Astrophys. J. Lett.",
    volume = "951",
    number = "1",
    pages = "L11",
    year = "2023",
    note = "[Erratum: Astrophys.J.Lett. 971, L27 (2024), Erratum: Astrophys.J. 971, L27 (2024)]"
}

@article{Miles:2024seg,
    author = "Miles, Matthew T. and others",
    title = "{The MeerKAT Pulsar Timing Array: the first search for gravitational waves with the MeerKAT radio telescope}",
    eprint = "2412.01153",
    archivePrefix = "arXiv",
    primaryClass = "astro-ph.HE",
    doi = "10.1093/mnras/stae2571",
    journal = "Mon. Not. Roy. Astron. Soc.",
    volume = "536",
    number = "2",
    pages = "1489--1500",
    year = "2024"
}

@article{Kajantie:1996mn,
    author = "Kajantie, K. and Laine, M. and Rummukainen, K. and Shaposhnikov, Mikhail E.",
    title = "{Is there a~ hot electroweak phase transition at $m_H \gtrsim m_W$?}",
    eprint = "hep-ph/9605288",
    archivePrefix = "arXiv",
    reportNumber = "CERN-TH-96-126, HD-THEP-96-15, IUHET-333",
    doi = "10.1103/PhysRevLett.77.2887",
    journal = "Phys. Rev. Lett.",
    volume = "77",
    pages = "2887--2890",
    year = "1996"
}

@article{Aoki:2006we,
    author = "Aoki, Y. and Endrodi, G. and Fodor, Z. and Katz, S. D. and Szabo, K. K.",
    title = "{The Order of the quantum chromodynamics transition predicted by the standard model of particle physics}",
    eprint = "hep-lat/0611014",
    archivePrefix = "arXiv",
    doi = "10.1038/nature05120",
    journal = "Nature",
    volume = "443",
    pages = "675--678",
    year = "2006"
}

@article{Kamionkowski:1993fg,
    author = "Kamionkowski, Marc and Kosowsky, Arthur and Turner, Michael S.",
    title = "{Gravitational radiation from first order phase transitions}",
    eprint = "astro-ph/9310044",
    archivePrefix = "arXiv",
    reportNumber = "IASSNS-HEP-93-44, FERMILAB-PUB-93-235-A",
    doi = "10.1103/PhysRevD.49.2837",
    journal = "Phys. Rev. D",
    volume = "49",
    pages = "2837--2851",
    year = "1994"
}

@article{Dolan:1973qd,
    author = "Dolan, L. and Jackiw, R.",
    title = "{Symmetry Behavior at Finite Temperature}",
    reportNumber = "MIT-CTP-406",
    doi = "10.1103/PhysRevD.9.3320",
    journal = "Phys. Rev. D",
    volume = "9",
    pages = "3320--3341",
    year = "1974"
}

@article{Weinberg:1974hy,
    author = "Weinberg, Steven",
    title = "{Gauge and Global Symmetries at High Temperature}",
    reportNumber = "PRINT-74-0689 (HARVARD)",
    doi = "10.1103/PhysRevD.9.3357",
    journal = "Phys. Rev. D",
    volume = "9",
    pages = "3357--3378",
    year = "1974"
}

@article{Linde:1980ts,
    author = "Linde, Andrei D.",
    title = "{Infrared Problem in Thermodynamics of the Yang-Mills Gas}",
    reportNumber = "LEBEDEV-80-106",
    doi = "10.1016/0370-2693(80)90769-8",
    journal = "Phys. Lett. B",
    volume = "96",
    pages = "289--292",
    year = "1980"
}

@article{Linde:1978px,
    author = "Linde, Andrei D.",
    title = "{Phase Transitions in Gauge Theories and Cosmology}",
    reportNumber = "LEBEDEV-78-166",
    doi = "10.1088/0034-4885/42/3/001",
    journal = "Rept. Prog. Phys.",
    volume = "42",
    pages = "389",
    year = "1979"
}

@article{Parwani:1991gq,
    author = "Parwani, Rajesh R.",
    title = "{Resummation in a hot scalar field theory}",
    eprint = "hep-ph/9204216",
    archivePrefix = "arXiv",
    reportNumber = "ITP-SB-91-64",
    doi = "10.1103/PhysRevD.45.4695",
    journal = "Phys. Rev. D",
    volume = "45",
    pages = "4695",
    year = "1992",
    note = "[Erratum: Phys.Rev.D 48, 5965 (1993)]"
}

@article{Farakos:1994kx,
    author = "Farakos, K. and Kajantie, K. and Rummukainen, K. and Shaposhnikov, Mikhail E.",
    title = "{3-D physics and the electroweak phase transition: Perturbation theory}",
    eprint = "hep-ph/9404201",
    archivePrefix = "arXiv",
    reportNumber = "CERN-TH-6973-94, IUHET-273",
    doi = "10.1016/0550-3213(94)90173-2",
    journal = "Nucl. Phys. B",
    volume = "425",
    pages = "67--109",
    year = "1994"
}

@article{Braaten:1995cm,
    author = "Braaten, Eric and Nieto, Agustin",
    title = "{Effective field theory approach to high temperature thermodynamics}",
    eprint = "hep-ph/9501375",
    archivePrefix = "arXiv",
    reportNumber = "NUHEP-TH-95-2",
    doi = "10.1103/PhysRevD.51.6990",
    journal = "Phys. Rev. D",
    volume = "51",
    pages = "6990--7006",
    year = "1995"
}

@article{Kajantie:1995dw,
    author = "Kajantie, K. and Laine, M. and Rummukainen, K. and Shaposhnikov, Mikhail E.",
    title = "{Generic rules for high temperature dimensional reduction and their application to the standard model}",
    eprint = "hep-ph/9508379",
    archivePrefix = "arXiv",
    reportNumber = "CERN-TH-95-226, HU-TFT-95-50, IUHET-312",
    doi = "10.1016/0550-3213(95)00549-8",
    journal = "Nucl. Phys. B",
    volume = "458",
    pages = "90--136",
    year = "1996"
}

@article{Lofgren:2023sep,
    author = {L{\"o}fgren, Johan},
    title = "{Stop comparing resummation methods}",
    eprint = "2301.05197",
    archivePrefix = "arXiv",
    primaryClass = "hep-ph",
    doi = "10.1088/1361-6471/ad074b",
    journal = "J. Phys. G",
    volume = "50",
    number = "12",
    pages = "125008",
    year = "2023"
}

@article{Costa:2025pew,
    author = "Costa, Francesco and Hoefken Zink, Jaime and Lucente, Michele and Pascoli, Silvia and Rosauro-Alcaraz, Salvador",
    title = "{ELENA: a software for fast and precise computation of first order phase transitions and gravitational waves production in particle physics models}",
    eprint = "2510.00289",
    archivePrefix = "arXiv",
    primaryClass = "hep-ph",
    month = "9",
    year = "2025"
}

@inproceedings{Wang:2025wht,
    author = "Wang, Mengshen and Zhang, Zuocheng and Xu, Hua",
    title = "{Discriminating Between Models of the Nanohertz Gravitational-Wave Background with Pulsar Timing Arrays}",
    eprint = "2510.22713",
    archivePrefix = "arXiv",
    primaryClass = "astro-ph.CO",
    month = "10",
    year = "2025"
}

@article{Goncharov:2024htb,
    author = "Goncharov, Boris and others",
    title = "{Reading signatures of supermassive binary black holes in pulsar timing array observations}",
    eprint = "2409.03627",
    archivePrefix = "arXiv",
    primaryClass = "astro-ph.HE",
    doi = "10.1038/s41467-025-65450-3",
    journal = "Nature Commun.",
    volume = "16",
    number = "1",
    pages = "9692",
    year = "2025"
}

@article{Antel:2023hkf,
    author = "Antel, C. and others",
    title = "{Feebly-interacting particles: FIPs 2022 Workshop Report}",
    eprint = "2305.01715",
    archivePrefix = "arXiv",
    primaryClass = "hep-ph",
    reportNumber = "CERN-TH-2023-061, DESY-23-050, FERMILAB-PUB-23-149-PPD, INFN-23-14-LNF, JLAB-PHY-23-3789, LA-UR-23-21432, MITP-23-015",
    doi = "10.1140/epjc/s10052-023-12168-5",
    journal = "Eur. Phys. J. C",
    volume = "83",
    number = "12",
    pages = "1122",
    year = "2023"
}

@article{Agrawal:2021dbo,
    author = "Agrawal, Prateek and others",
    title = "{Feebly-interacting particles: FIPs 2020 workshop report}",
    eprint = "2102.12143",
    archivePrefix = "arXiv",
    primaryClass = "hep-ph",
    doi = "10.1140/epjc/s10052-021-09703-7",
    journal = "Eur. Phys. J. C",
    volume = "81",
    number = "11",
    pages = "1015",
    year = "2021"
}

@article{Ekstedt:2024etx,
    author = "Ekstedt, Andreas and Schicho, Philipp and Tenkanen, Tuomas V. I.",
    title = "{Cosmological phase transitions at three loops: The final verdict on perturbation theory}",
    eprint = "2405.18349",
    archivePrefix = "arXiv",
    primaryClass = "hep-ph",
    reportNumber = "HIP-2024-15/TH",
    doi = "10.1103/PhysRevD.110.096006",
    journal = "Phys. Rev. D",
    volume = "110",
    number = "9",
    pages = "096006",
    year = "2024"
}

@article{Gould:2023ovu,
    author = "Gould, Oliver and Tenkanen, Tuomas V. I.",
    title = "{Perturbative effective field theory expansions for cosmological phase transitions}",
    eprint = "2309.01672",
    archivePrefix = "arXiv",
    primaryClass = "hep-ph",
    reportNumber = "NORDITA 2023-037",
    doi = "10.1007/JHEP01(2024)048",
    journal = "JHEP",
    volume = "01",
    pages = "048",
    year = "2024"
}

@article{Athron:2022jyi,
    author = "Athron, Peter and Balazs, Csaba and Fowlie, Andrew and Morris, Lachlan and White, Graham and Zhang, Yang",
    title = "{How arbitrary are perturbative calculations of the electroweak phase transition?}",
    eprint = "2208.01319",
    archivePrefix = "arXiv",
    primaryClass = "hep-ph",
    doi = "10.1007/JHEP01(2023)050",
    journal = "JHEP",
    volume = "01",
    pages = "050",
    year = "2023"
}

@article{Schicho:2021gca,
    author = {Schicho, Philipp M. and Tenkanen, Tuomas V. I. and {\"O}sterman, Juuso},
    title = "{Robust approach to thermal resummation: Standard Model meets a singlet}",
    eprint = "2102.11145",
    archivePrefix = "arXiv",
    primaryClass = "hep-ph",
    doi = "10.1007/JHEP06(2021)130",
    journal = "JHEP",
    volume = "06",
    pages = "130",
    year = "2021"
}

@article{Schicho:2022wty,
    author = "Schicho, Philipp and Tenkanen, Tuomas V. I. and White, Graham",
    title = "{Combining thermal resummation and gauge invariance for electroweak phase transition}",
    eprint = "2203.04284",
    archivePrefix = "arXiv",
    primaryClass = "hep-ph",
    reportNumber = "HIP-2022-2/TH, NORDITA 2022-009",
    doi = "10.1007/JHEP11(2022)047",
    journal = "JHEP",
    volume = "11",
    pages = "047",
    year = "2022"
}

@article{Li:2023bxy,
    author = "Li, Shao-Ping and Xie, Ke-Pan",
    title = "{Collider test of nano-Hertz gravitational waves from pulsar timing arrays}",
    eprint = "2307.01086",
    archivePrefix = "arXiv",
    primaryClass = "hep-ph",
    doi = "10.1103/PhysRevD.108.055018",
    journal = "Phys. Rev. D",
    volume = "108",
    number = "5",
    pages = "055018",
    year = "2023"
}

\end{document}